# Active Force Dynamics in Red Blood Cells Under Non-Invasive Optical Tweezers

Arnau Dorn[1,2], Clara Luque-Rioja[1,2], Macarena Calero[2], Diego Herráez-Aguilar[3], Francisco Monroy[1,2, #] and Niccolò Caselli[1,2,*]

[1] *Departamento de Química Física, Universidad Complutense de Madrid, Ciudad Universitaria s/n, 28040 Madrid, Spain.*
[2] *Translational Biophysics, Instituto de Investigación Sanitaria Hospital Doce de Octubre, 28041 Madrid, Spain.*
[3] *Instituto de Investigaciones Biosanitarias, Universidad Francisco de Vitoria, Ctra. Pozuelo-Majadahonda, Pozuelo de Alarcón, Madrid, Spain.*

**Email:** *ncaselli@ucm.es; #monroy@ucm.es
**Keywords:** Optical tweezers; Red blood cell; Mechanobiology.

**Abstract**

Red blood cells (RBCs) sustain mechanical stresses associated with microcirculatory flow through ATP-driven plasma membrane flickering. This is an active phenomenon driven by motor proteins that regulate interactions between the spectrin cytoskeleton and the lipid bilayer; it is manifested in RBC shape fluctuations reflecting the cell's mechanical and metabolic state. Yet, direct quantification of the forces and energetic costs underlying this non-equilibrium behavior remains challenging due to the invasiveness of existing techniques. Here, a minimally invasive method that combines bead-free, low-power optical tweezers with high-speed video microscopy is employed to track local membrane forces and displacements in single RBCs during the same time window. This independent dual-channel measurement enables the construction of a mechano-dynamic phase space for RBCs under different chemical treatments, that allows for differentiating between metabolic and structural states based on their fluctuation-force signatures. Quantification of mechanical work during flickering demonstrates that membrane softening enhances fluctuations while elevating energy dissipation. The proposed optical tweezers methodology provides a robust framework for mapping the active mechanics of living cells, enabling precise probing of cellular physiology and detection of biomechanical dysfunction in diseases.

**Introduction**

Human red blood cells (RBCs) are essential for oxygen delivery and gas exchange, relying on their extraordinary deformability to squeeze through narrow capillaries [1]. This mechanical resilience is not passive since it reflects a dynamic balance between cell structure and metabolic activity [1,2]. At the core of this adaptability lies the RBC membrane, which is a fluid lipid bilayer supported by a skeleton that provides elasticity and mechanobiological stability [3–5]. Such underlying rigid structure consists of a two-dimensional network of spectrin filaments interconnecting junctional complexes based on short and dynamic actin filaments [6,7]. RBC membrane exhibits spontaneous nanometer-scale fluctuations, termed *flickering* [8], which are powered by ATP and act as sensitive indicator of the cell's mechanical integrity and physiological state [9–11]. Flickering is an out-of-equilibrium phenomenon driven by the cell's metabolism, that reflects ongoing remodeling activities within the membrane-cytoskeleton complex [7,12,13]. Recent studies even suggested direct force contributions from conformational changes within the spectrin network [14]. More profound alterations in RBC shape or deformability are linked to cellular aging and pathological states, including hemolytic anemia and metabolic disorders, underscoring the diagnostic value of understanding RBC biomechanics [15,16]. However, despite the potential use of RBC flickering for cellular diagnostics, directly linking these fluctuations to local mechanical forces and energy dissipation has remained experimentally elusive. Most available techniques strongly perturb the cell or lack the spatial and temporal resolution needed to capture native dynamics. Atomic force microscopy provides detailed stiffness maps, but requires direct contact, hence altering cell behavior [17]. Microfluidic platforms impose strong mechanical constraints that can obscure intrinsic viscoelastic responses [18]; while fluorescence-based traction force microscopy often relies on genetic or chemical modifications that may interfere with cellular function [19]. Optical tweezers (OTs) offer sub-nanometric precision and microsecond resolution and have been used to measure RBC membrane fluctuations [9,12]. However, traditional implementations rely on beads chemically attached to the membrane, introducing artifacts and complicating force calibration procedure. Bead-free OTs applied directly to entire RBCs can circumvent these issues by exploiting the cell's intrinsic refractive index contrast [20–25]. Bead-free OTs approaches typically apply the trapping force either to the whole cell or to two opposite membrane positions (for stretching/squeezing), thus lacking the necessary spatial resolution to probe arbitrary membrane locations. Moreover, these methods usually detect either RBCs deformation or force, but not both independently during the same observational time.

Here, we overcome these limitations by combining low-power direct optical tweezers with high-speed imaging to simultaneously measure local forces and flickering amplitudes on single RBCs under physiological and mechano-biologically perturbed conditions. Optical traps were directly applied to selected local membrane positions without intermediary beads, allowing to measure local forces through laser momentum changes, as the OTs laser light interacts with the cell membrane [26]. This approach bypasses the limitations of traditional OTs setups, which require bead attachment, assume harmonic trap behavior, and depend on precise calibration tied to bead geometry and medium viscosity [27,28]. Instead, momentum-based force detection enables accurate, real-time measurements even under large displacements or nonlinear responses. Crucially, the use of low-power IR lasers ensures that intrinsic membrane dynamics remain unperturbed, allowing reliable access to native mechanical behavior across different physiological and perturbed states. Our approach enabled simultaneous, independent measurement of local membrane flickering and active force generation at multiple positions on single RBCs. We applied this method to healthy RBCs and to RBCs subjected to targeted perturbations that mimic pathological conditions, including cytoskeletal disruption via Latrunculin A (LatA) [29,30], ATP depletion via iodoacetamide and inosine [9,10], and chemical fixation via glutaraldehyde (GA) [31]. These treatments may model key features of early stage diseases such as hereditary spherocytosis, elliptocytosis, metabolic disorders, and oxidative stress-related anemias [1]. By comparing the native and perturbed mechanobiological states, we revealed how cytoskeletal integrity and metabolic activity shape membrane behavior and quantified the mechanical work and energy dissipation associated with flickering. We mapped the relationship between membrane fluctuations and mechanical forces, defining a mechano-dynamic signature of the cell's structural and metabolic state. This approach aims at providing a new, minimally invasive framework for probing RBCs mechanobiology and detecting early mechanical dysfunction. The cell's flickering mechanics can be considered as a quantitative mechano-dynamic signature that precedes morphological changes. Since flickering arises from ATP-dependent forces and cytoskeletal coupling, subtle metabolic or cytoskeletal impairments shift the joint force-displacement statistics before any remarkable shape alteration occurs. Consequently, this sensitive, non-invasive analysis of flickering mechanics can serve as an early biomarker for RBC dysfunction and subclinical hematological disorders.

**Results**

**Local membrane fluctuations and forces**

To study the dynamics of *in vitro* single RBCs we prepared the samples by purifying the extracted blood from a single male healthy donor, suspending the cells in a phosphate saline buffer with glucose, and placing them into a sealed observation chamber (see Methods and Supplementary Figure S1). We focused on RBCs exhibiting a discocyte shape, as they represented physiologically healthy cells [32] and constituted the majority of the cells in the sample. RBCs settled by gravity onto the bottom chamber's glass coverslip, with the inner concave surface of their lower discocyte rim contacting the glass. This basal adhesion ensured that the outer equatorial cell rim remained free to oscillate during the measurements of membrane dynamics, as represented in Figures 1 a) and S1. The flickering of individual RBCs, corresponding to membrane equatorial fluctuations, was analyzed by monitoring the membrane contour over time using high-speed video microscopy (see Methods). By tracking the membrane position for different frames as a function of time, we evaluated the contour radius $R(\theta,t)$ at angle $\theta$ and time $t$. These radial coordinates are defined in Figure 1 b). The information extracted from tracking was used to estimate the local fluctuations, $\delta h$, represented as green arrow in Figure 1b) and defined as the radial displacement relative to the time averaged position $\delta h(\theta,t) = R(\theta,t) - \langle R(\theta,t) \rangle$; where $\langle\ \rangle$ represents the average over the total observational time. During video acquisitions, a pattern of eight optical traps was precisely positioned on the equatorial rim of the RBC, as illustrated in Figure 1 a)-b). This configuration ensured overall cell stability and minimized mechanical cross-talks that could arise from the propagation of viscoelastic deformation between rapidly activated neighboring traps positions.

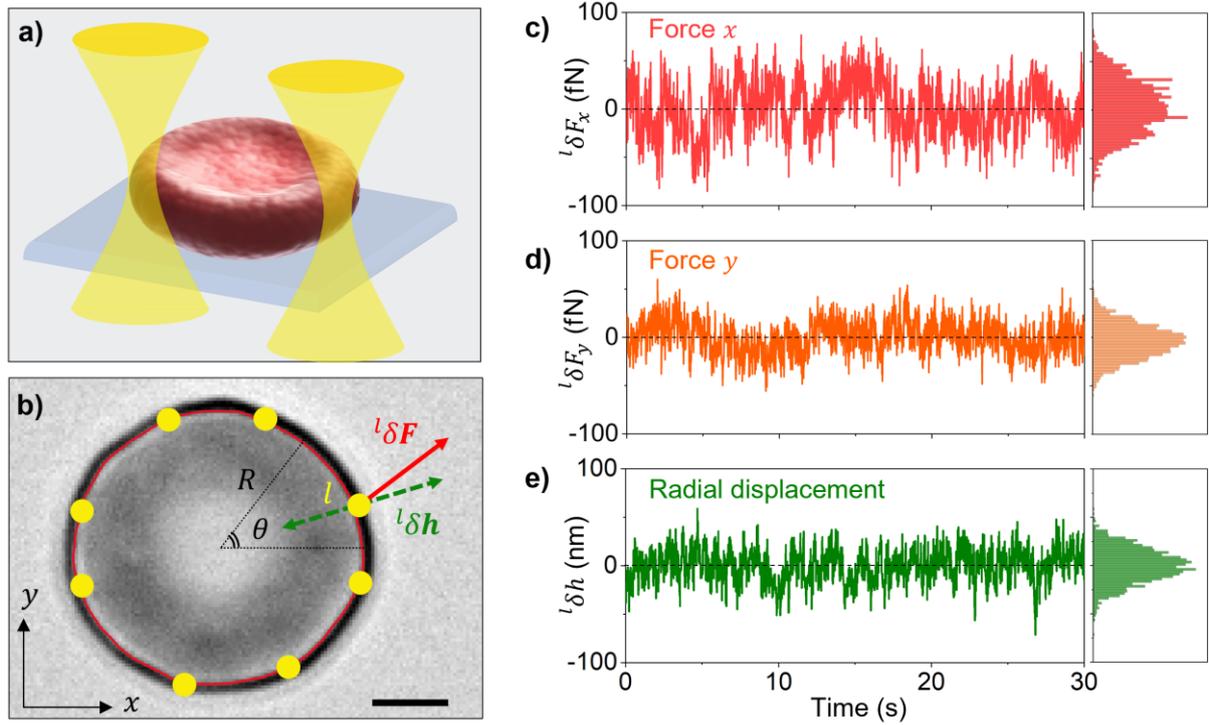

**Figure 1. Simultaneous detection of local deformations and forces generated by RBC membrane flickering.** **a)** Schematics of the multi-trap optical tweezers (OTs, represented as focused Gaussian laser beams) directly applied to RBC in contact with glass substrate. **b)** Bright-field microscopy image of a discocyte RBC observed in the equatorial $xy$ plane. The radial coordinates are $R$, $\theta$. Scale bar is 2 µm. The red curve represents the membrane contour and yellow dots identify the positions where 8 optical traps were deployed. Green arrows represent the local radial displacement $^l\delta\boldsymbol{h}$ measured by video microscopy at position of trap $l$ with respect to the membrane mean position. For the sake of clarity, $^l\delta\boldsymbol{h}$ amplitude was not drawn to scale. Red arrow shows the local force variation $^l\delta\boldsymbol{F}$ with respect to the mean value, measured by OTs at position of trap $l$. **c)-e)** Temporal series of force component $^l\delta F_x$, $^l\delta F_y$, and radial displacement $^l\delta h$ along with the corresponding distributions (right panels). The laser power employed was 1.5 mW/trap.

The multi-trap OTs were induced by a CW-IR laser emitting at $\lambda = 1064$ nm in combination with acousto optic deflectors and a high numerical aperture objective that allowed us to generate the desired number of focused spots on the sample plane (with lateral size limited by diffraction), reported as yellow dots in Figure 1 b). The schematics of the experimental setup was reported in Supplementary Figure S2. The OTs traps can be precisely positioned at desired spots by means of the integrated software (SENSOCELL, by IMPETUX). For every trap (identified by index $l$) placed at the RBC membrane rim, the OTs force sensor acquired the temporal series of the in-plane $(x, y)$ components of the force vector $^l\boldsymbol{F}^{trap}$ that corresponded to the force induced by the optical trap $l$. The local force exerted by the RBC membrane, was determined as: $^l\boldsymbol{F} = -\left(\,^l\boldsymbol{F}^{trap} - \,^l\boldsymbol{F}^{off}\right)$, by applying newton's third law, where $^l\boldsymbol{F}^{off}$ is the baseline force detected when the studied RBC was moved away from the trapping region. This procedure ensured that any offset in the baseline force measurement due to the absolute position of the traps within the field of view was eliminated. The value of the modulus of the force $^l\boldsymbol{F}$ was in the range of 0.1-1.0 pN for OTs operating at a laser power of 1.5 mW/trap ($P_1$). This small force magnitude ensured that the applied trapping was weak, yet accurately measurable with the experimental resolution of 0.02 pN. Nevertheless, to characterize the active membrane processes our analysis was focused on the dynamics of the local force fluctuations, not on the absolute magnitude of the applied force.

Figures 1c) and 1d) present the time series for both measured force components generated by a single optical trap at location $l$ on a healthy RBC membrane. They were evaluated with respect to their mean values over the total observation time: $^l\delta F_i = {}^lF_i - \langle {}^lF_i \rangle$ for $i = x, y$. In Figure 1 e) we reported the local radial displacement $^l\delta h$ corresponding to the fluctuations of the membrane position where the trap was generated. Every time series was plotted alongside their probability distributions (right panels), which highlight the stochastic nature of these dynamic variables. To estimate the local trap stiffness ($K$), we applied a single optical trap at power $P_1$ on healthy RBCs and measured the force $\delta F_i$ as a function of a controlled displacement $\delta s_i$ in both directions, for different positions along the membrane contour (see Supplementary Figure S3). The stiffness was calculated as the slope of the force-displacement curve in the linear regime, approximately in the range (-0.5, 0.5) $\mu$m around the initial position of the trap, assuming the relationship $\delta F_i = -K_i \delta s_i$ holds for both directions. The trap stiffness was dependent on the local spatial distribution of the refractive index, which was governed by the specific geometry of the trapped cell. Since the local gradient of the refractive index was maximized across the radial direction of the membrane equatorial border (crossing the cytoplasm-buffer interface), the optical gradient force detected was larger in this direction. Therefore, the radial component of the stiffness was consistently larger ($K_R \approx 2.4$ pN/$\mu$m) with respect to the tangential one ($K_T \approx 0.6$ pN/$\mu$m), as reported in Supplementary Figure S4 and Table T1.

**Flickering of free-standing RBCs**

We characterized the membrane fluctuations of free-standing RBCs, i.e., cells deposited on glass coverslip in the absence of optical trapping, for various perturbations that affected cell structure and metabolism. We studied four samples: healthy RBCs and cells subjected to three previously described chemical treatments designed to reduce membrane rigidity (LatA), suppress cell energy reservoir (ATP depletion), or increase overall cell rigidity (fixation). Even under such treatments, we studied only RBCs with discocyte shape, which remained predominant across all conditions, ensuring that the measurements reflected the biomechanical properties of physiologically relevant cells. To quantify flickering amplitude, the standard deviation, $\sigma_{\delta h}$, of radial fluctuations obtained by video microscopy, was used as an effective indicator [23,33]. Figure 2a) shows flickering maps where the local values of $\sigma_{\delta h}$, calculated for the studied cell over the full time series (duration 30 s), were plotted along the average membrane contour for four representative RBCs, each from one of the four conditions. These maps were displayed using the same color scale to facilitate direct comparison and clearly reveal the distinct patterns. Additional flickering maps for RBCs of the four samples can be found in Supplementary Figures S5-S8. RBCs with membrane softened by LatA showed larger fluctuation and stronger flickering; healthy RBCs exhibited intermediate flickering amplitude; ATP depleted cells showed reduced flickering and fixed RBCs presented very small flickering and cold flickering maps. In the latter case, cellular activity was completely suppressed and the residual membrane motions arose solely from passive, thermally driven fluctuations that were confined by the intact cytoskeletal structure. Moreover, $\sigma_{\delta h}$ maps revealed spatially heterogeneous activity, with local hot spots, especially in healthy and LatA-treated RBCs, whereas ATP-depleted and fixed cells exhibited a more uniform distribution. To assess global statistical differences among the studied samples, we evaluated the probability density function (PDF) of radial fluctuation, derived from $\delta h$ events occurring within a given amplitude range across all observation times, membrane locations and cells from the given sample. Figure 2 b) shows the PDF normalized to the total integral for each sample. The largest fluctuation amplitudes and the broadest PDF corresponded to LatA-treated sample with reduced membrane rigidity. Healthy RBCs showed intermediate flickering amplitude, while ATP depleted cells exhibited reduced fluctuations and a narrower PDF. Fixation strongly reduced the flickering activity, resulting in constrained fluctuations and very narrow PDF. This trend was consistent with the quantitative comparison of the overall flickering magnitudes provided in Figure 2 c). Here, the distribution of the local values of $\sigma_{\delta h}$ for each sample confirmed the statistical significance of differences by a pairwise $t$-test at the level of significance set to 0.001 for all samples (see Methods). The statistical significance was even greater when each active sample was compared to the passive control (fixed RBCs), as reported in the detailed analysis of Supplementary Table T2. To evaluate the reproducibility and stability of local hot spots in the flickering maps, we performed subsequent time succession measurements on the same cells, as reported in Supplementary Figure S9. We found that the flickering distributions and the locations of these hot spots were persistent over time scales of few minutes. However, over longer observation periods (approximately 15 min), the hot spots shifted around the cell perimeter, and the overall fluctuation amplitude changed significantly. This temporal behavior highlights that the local fluctuation patterns were not fixed geometrical artifacts but rather a manifestation of the active, continuous reorganization of the underlying cytoskeletal network.

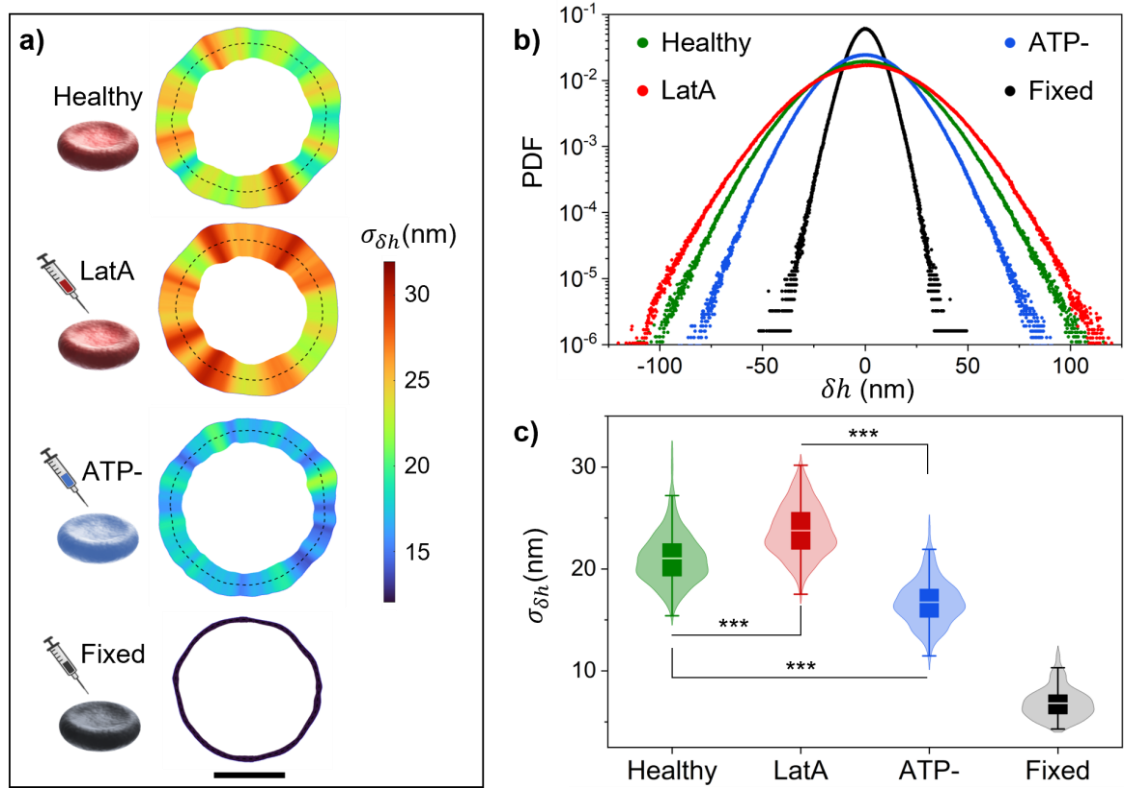

**Figure 2. Membrane flickering of free-standing cell under different treatments. a)** Flickering maps obtained in the absence of optical traps (free-standing cells) for healthy RBCs, RBCs treated with Latrunculin A (LatA), with iodoacetamide and inosine (ATP-) and fixed by glutaraldehyde. The maps were evaluated as the standard deviation, $\sigma_{\delta h}$, of the local radial membrane fluctuations $\delta h$. They are shown in a blue to red color scale in the range (12-32) nm. While the in-plane extension is visually exaggerated and out of scale, it accurately reflects the relative amplitude of the local fluctuations across all samples. Internal dashed lines are the membrane mean positions. Scale bar is 4 µm. **b)** Normalized probability density function (PDF) of the local fluctuation, $\delta h$, for all samples described in a). **c)** Distributions of the local values of $\sigma_{\delta h}$ for the observed samples. Internal boxes are the distributions standard deviations; horizontal transparent lines are the means. The error bars extend from the first quartile to the third quartile. The symbol *** indicates that the difference between the two labeled samples is statistically significant (pairwise two-sample $t$-test, $p$ <0.001). The definitions of individual symbols used in c) also apply to other similar figures in the manuscript. The distributions presented in b) and c) include data from $N$ =10 RBCs for the healthy, LatA, and ATP- samples, and $N$ =7 RBCs for the fixed sample.

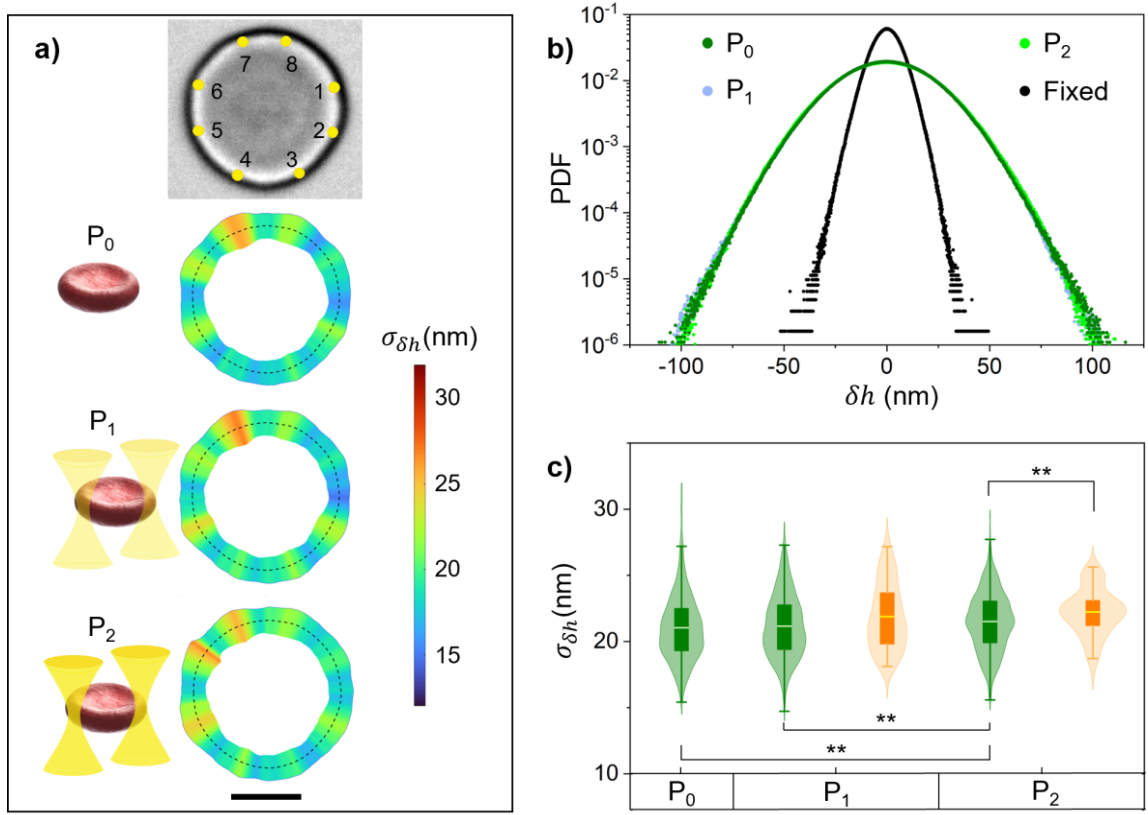

**Figure 3. Healthy RBCs flickering under optical tweezers. a)** Position of 8 optical traps placed on a healthy RBC membrane (yellow dots in upper panel). Flickering maps, $\sigma_{\delta h}$, for the same RBC at increasing OTs laser power (from second to bottom panel): $P_0$=0 mW, i.e., free-standing RBC, $P_1$=1.5 mW/trap, $P_2$=3.0 mW/trap. Scale bar is 4 μm. **b)** Normalized probability density function (PDF) of $\delta h$ evaluated over N=10 healthy RBCs for power equal to $P_0$ (dark green), $P_1$ (blue) and $P_2$ (light green); compared to N=7 fixed RBCs (black). **c)** Distribution of the local values of $\sigma_{\delta h}$ at each laser power, for the same cells as in **b)** and for all tracked membrane positions (green data), or only for positions where the optical traps were generated (orange data). The symbol ** indicates that the difference between the two labeled samples is statistically significant (pairwise two-sample $t$-test, $p$ <0.01).

**Optical trapping with minimal perturbation**

The foundation of our methodology lies in using optical tweezers as force sensors, rather than as tools for inducing cell immobilization or stretching [34]. To minimize perturbations to membrane flickering and prevent potential phototoxicity from prolonged exposure, we employed a low-power IR laser. To evaluate the effect of trapping power on membrane flickering, we compared the fluctuations across three conditions: free-standing RBCs ($P_0$=0 mW, without OTs) and the exact same individual cells under optical traps at laser power equal to $P_1$=1.5 mW/trap and $P_2$=3.0 mW/trap. Figure 3 a) shows flickering maps for a representative healthy RBC interacting with eight optical traps at laser power $P_0$, $P_1$ and $P_2$, respectively. The qualitative visual comparison indicates strong similarity between flickering maps at $P_0$ and $P_1$, with minor differences in amplitude distribution emerging at $P_2$. Quantitative analysis of PDF and $\sigma_{\delta h}$ distributions, calculated for flickering across N=10 healthy RBCs and all membrane locations at different trapping powers, are reported in Figure 3 b)-c), respectively. Detailed numerical analysis can be found in Table T3-T4. These data confirm that OTs operating at laser power $P_1$ did not cause significant changes in flickering amplitude when compared to free-standing cells. However, the higher power $P_2$ led to a slight modification of the $\sigma_{\delta h}$ distributions, that also occurred for the flickering measured at the trapped locations, represented by the orange data in Figure 3 c). We extended the flickering analysis at different OTs laser power to LatA-treated and ATP-depleted RBCs and representative $\sigma_{\delta h}$ maps and distributions for these conditions are shown in Figure 4. Detailed numerical analysis can be found in Supplementary Table T5-T6. For highly dynamic RBCs with reduced membrane rigidity (treated with LatA), the flickering amplitude was significantly suppressed under optical trapping, even for OTs working at the lowest laser power, $P_1$. However, the inhibition was not large enough to reach the level of healthy free-standing RBCs, which exhibited lower flickering amplitude than LatA-treated trapped cells. For RBCs with reduced metabolism (ATP-), flickering amplitude remained unaffected at laser power $P_1$, but augmented significantly at $P_2$. Moreover, for both treated samples presented in Figure 4, the flickering evaluated at the trap locations was significantly increased with respect to other membrane positions. The sole exception was observed in ATP-depleted RBCs trapped at power $P_2$. Furthermore, the trap stiffness assessed for the active states (healthy and LatA-treated RBCs) presented in this study remained unaffected by these specific chemical treatments, as reported in Supplementary Figure S4, providing a reliable reference for comparing force fluctuations. We observed a ~20% increase in fixed cells stiffness that could be explained by physical change due to protein crosslinking. To establish the optimal setting for effective force measurements, we chose OTs operating at $P_1$=1.5 mW/trap, since it was the lowest power needed to generate a stable local force (characterized by stiffness $K_R$~ 2.5 pN/μm) without altering the cell's global flickering dynamics. This is the reason why all the force measurements presented in the following sections were performed using eight traps at laser power $P_1$. To check the non-invasiveness of the chosen power level, we tested the reversibility of the flickering patterns by conducting subsequent time succession measurements on the same cells before, during, and after applying the OTs. As shown in Supplementary Figure S10, no statistically significant difference was found between the overall fluctuation distributions, demonstrating that the optical force induced only a slight and reversible mechanical perturbation without altering the flickering or causing permanent structural modifications to the RBCs.

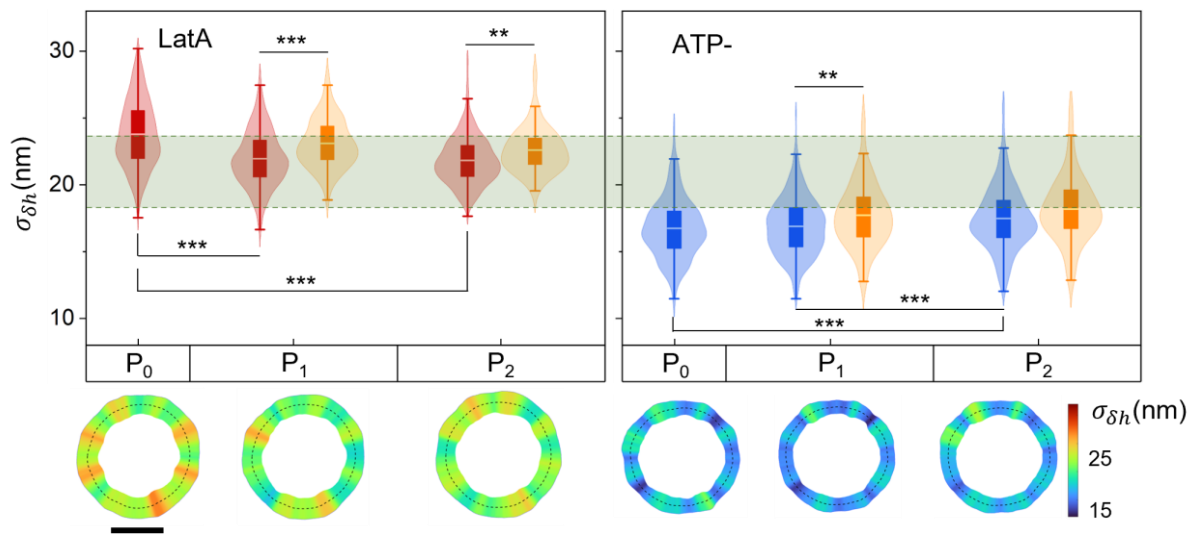

**Figure 4. Treated RBCs flickering under optical tweezers.** Distribution of the local values of $\sigma_{\delta h}$ when optical tweezers were applied on 8 positions of the cell membrane at laser power per trap equal to $P_0$=0 mW, $P_1$=1.5 mW, $P_2$=3.0 mW, respectively, for RBCs treated by Latrunculin A (LatA, left panel) and iodoacetamide/inosine (ATP-, right panel). $\sigma_{\delta h}$ distributions are evaluated for all sample cells and all the tracked membrane radial displacements (red boxes in the left panel; blue boxes in the right panel) as well as only at positions where the optical traps were activated (orange boxes). The symbol *** indicates $p$ <0.001 and ** indicates $p$ <0.01. The horizontal green stripe represents $\sigma_{\delta h}$ =(21.0 ± 2.5) nm for healthy RBCs at $P_1$, as the confidence interval defined by the mean ± standard deviation. Data from $N$ =10 RBCs for both, LatA, and ATP- samples. The lower panel reports typical RBCs flickering maps for a given cell at each laser power. Scale bar is 4 μm.

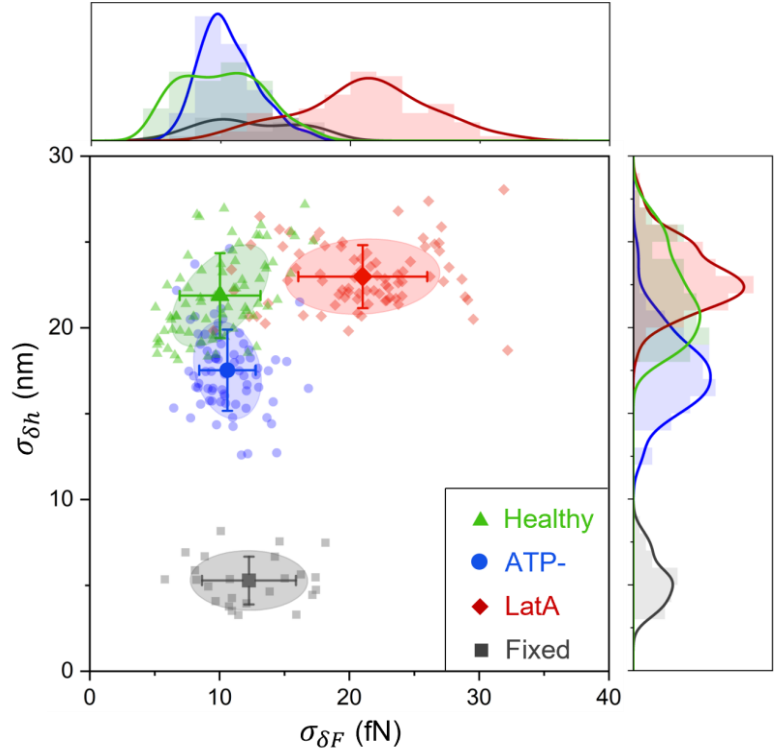

**Figure 5. RBC dynamics map.** Flickering and force fluctuations are reported in the $\sigma_{\delta h} - \sigma_{\delta F}$ space for every membrane position on which optical tweezers were applied and for different cell treatment. Each point in this space corresponds to the standard deviation of membrane radial fluctuations, $\sigma_{\delta h}$, and the standard deviation of radial component of optical trapping force, $\sigma_{\delta F}$, at a given membrane position. Different groups correspond to healthy RBCs (green triangles); RBCs treated with Latrunculin A (LatA, red diamonds); RBCs treated with inosine and iodoacetamide (ATP-, blue circles); RBCs treated with Glutaraldehyde (Fixed, gray squares). The bigger symbol in each ensemble is the mean value; error bars are the standard deviation of the group distribution. Ellipses correspond to the 70% confidence regions. The top and right distributions are the projections of the dynamics map on the $\sigma_{\delta F}$ and $\sigma_{\delta h}$ axis, respectively. Data are reported for OTs at laser power equal to $P_1$=1.5 mW/trap and were extracted form $N$ =10 RBCs for the healthy, LatA, and ATP- samples, and $N$ =7 RBCs for the fixed sample.

**Mapping the force-displacement variance space**

We explored the relationship between flickering and force fluctuations by performing simultaneous measurements of local membrane displacements (via video-microscopy) and local trapping forces (via OTs operating at the previously validated low power $P_1$) at eight positions across the RBCs membrane. For each trap positions, we calculated both standard deviation of radial displacement, $\sigma_{\delta h}$, and standard deviation of radial component of force fluctuation, $\sigma_{\delta F}$. The radial force component ($^l F_R$) was derived from the measured in-plane force components ($^l F_x$, $^l F_y$) by projecting them onto the local radial vector at the trap position, as detailed in the Methods section. The radial force fluctuation ($^l \delta F_R$) was then calculated by subtracting the time-averaged radial force ($\langle F_R \rangle$) from the instantaneous radial force values. Plotting the two parameters $\sigma_{\delta h}$ and $\sigma_{\delta F}$ against each other generated a two-dimensional dynamic map, presented in Figure 5. This map shows distinct clusters of data points for each of the four studied RBC samples. The mean values of $\sigma_{\delta h}$ and $\sigma_{\delta F}$ were reported with the associated uncertainty and statistical significance in Supplementary Tables T7 and T8, respectively. Healthy RBCs (green triangles) exhibited high flickering amplitude and moderate force variance. LatA-treated cells (red diamonds), with disrupted actin filaments that reduced the membrane rigidity, occupied a region with significantly higher $\sigma_{\delta h}$, and $\sigma_{\delta F}$, with respect to the healthy sample, hence indicating more dynamic membrane movements. ATP-depleted cells (blue circles), lacking metabolic energy for active processes, clustered at lower values of both $\sigma_{\delta h}$ and $\sigma_{\delta F}$. Finally, fixed cells (grey squares), representing a passive system, showed minimal fluctuations of both displacement and force. The clear separation between these samples, highlighted by 70% confidence ellipses and by distributions projected on the coordinate axes, demonstrates that the combined analysis of fluctuation amplitude and force variation provides a robust quantitative signature capable of distinguishing different RBC mechanobiological states. A dynamic map for the same set of RBCs at higher laser power, P$_2$, was shown in Supplementary Figure S11 and analyzed in Table T7, revealing minimal differences compared to the map at power P$_1$ (Figure 5). At higher power, force variations $\sigma_{\delta F}$ were slightly larger across all samples, while spatial fluctuations $\sigma_{\delta h}$ generally showed no significant variation. The exception to this trend was the sample of LatA-treated RBCs, which exhibited similar $\sigma_{\delta h}$ with respect to healthy cells at power P$_2$. This behavior represents a decrease in the membrane fluctuations for LatA-treated RBCs when interacting with OTs, as highlighted also in Figure 4. To evaluate the linear correlation of the dynamic data mapped for each RBCs sample, we calculated the Pearson coefficient ($P$), as reported in Supplementary Figure S12. This parameter is an estimator of the linear correlation between two variables, in this case $\sigma_{\delta h}$ and $\sigma_{\delta F}$ [35]. A high correlation ($P > 0.5$) was found only in healthy RBCs, whereas all the treatments applied in this study nearly eliminated the correlation, at both laser powers P$_1$ and P$_2$.

**Power dissipated during flickering**

The ability to measure both force and displacement during the same time frame allowed us to estimate the mechanical work performed by the membrane against the optical trap, providing direct insight into the energetics underlying the observed dynamics. For each trapping site, $l$ , the time-resolved accumulated work was computed in the radial direction using the discrete Stratonovich formula [36,37]:

$$W(t) = \int_0^t F_R(t') \circ dR(t') \approx \sum_{m=0}^{n-1} \bar{F}_R^{m+1/2} \cdot [R(t_{m+1}) - R(t_m)],$$

The mid-point force $\bar{F}_R^{m+1/2}$ was obtained by the time averaging of the values of the radial force at the given trapping site, acquired at 3125 Hz sampling rate, over the integration time of a single video frame defined as $\Delta t_m = t_{m+1} - t_m$, which corresponds to the 65 Hz frame rate used for displacement tracking (see Methods). This Stratonovich discretization symmetrically evaluates the product of force and displacement increments over finite sampling windows. The cumulative work $W(t)$ thus represents the total energy exchanged between the membrane and the optical trap up to time $t$. Figure 6 a) presents the work accumulated as a function of time, averaged over all trap positions ($l$) and over $N$ RBCs, for each sample. The average curves clearly establish a hierarchy: the largest work accumulation during flickering was produced by the softened membranes (RBCs treated with LatA), followed in succession by healthy and ATP-depleted cells. Minimal work was produced by the rigidified membranes of fixed RBCs. To study the statistical differences among the four samples, we analyzed the distributions of dynamic measurements, specifically the dissipated power, $P$, at every membrane position where optical traps were generated. After evaluating $W(t)$ for every trap, we extracted the power dissipated by the membrane flickering through mechanical fluctuations against the optical trap during the total observational time $\Delta t = 30$ s, as $P = W(\Delta t)/\Delta t$. The resulting distribution of dissipated power for each sample was reported in Figure 6 b). Detailed numerical analysis can be found in Table T9. A clear hierarchy also emerged in the power distributions, which was consistent with the activity levels observed in Figures 2 and 5. LatA-treated RBCs displayed the highest rate of work accumulation and dissipated significantly more power than all other groups (see $t$-test in Table T9), thus correlating their increased fluctuation amplitude and force variance with a higher energetic cost. Healthy RBCs exhibited an intermediate power dissipation level. ATP-depleted cells dissipated significantly less power with respect to the healthy sample. These findings reflect the reduced metabolic and mechanical activity of RBCs treated with iodoacetamide and inosine. Consistent with their passive nature, fixed RBCs exhibited the lowest power dissipation, showing a statistically significant difference from the other three groups.

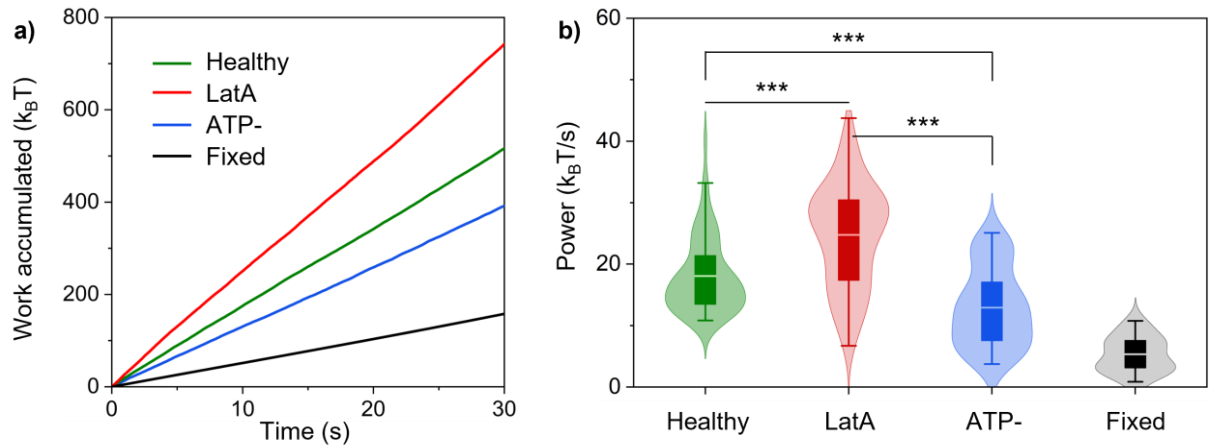

**Figure 6. Work accumulated and power dissipated during flickering. a)** Work accumulated in units of $k_B T$ as a function of time, averaged over all the observed membrane positions and RBCs of a given sample, at trapping power equal to P$_1$=1.5 mW/trap. **b)** Dissipated power in units of $k_B T/s$ for each RBC sample at P$_1$. The symbol *** indicates $p<0.001$. Data were extracted from $N=10$ RBCs for the healthy, LatA, and ATP- samples, and $N=7$ RBCs for the fixed sample.

**Discussion**

We developed a method to study single RBCs mechanics by measuring, during the same observational time, local membrane flickering and force exerted at a position defined by the center of the optical trap. On the one hand, flickering was associated with in-plane radial fluctuations of the membrane contour, which was detected by high-speed video microscopy. On the other hand, low-power optical tweezers were directly applied to RBCs membrane to retrieve the local RBCs force without altering the intrinsic flickering or the cell shape. The evaluation of the trap stiffness (about 2.5-3.0 pN/μm) proved that RBCs local membrane direct trapping was effective, ensuring that OTs can be used as local force sensors. We tested RBCs under different chemical treatments to benchmark activity and rigidity states of the cells. Since we found that trap stiffness and RBCs discocyte shape were independent for the active states (healthy and LatA), we inferred that changing the membrane rigidity or the cell metabolic activity did not affect the refractive index distribution of the membrane, in agreement with previous studies [38]. The ~20% stiffness increase observed in fixed cells, consistent with protein crosslinking, suggested a marginal alteration in the refractive index distribution. Critically, this change remained within the same order of magnitude across all conditions and did not affect the primary objective of our work, which was the analysis of membrane displacement and force fluctuations.

The flickering analysis on free-standing RBCs, without optical trapping, revealed statistically significant differences in membrane fluctuations when the cells were subject to chemical treatments, as reported in Figure 2. The highest and most heterogeneous flickering activity was observed for RBCs treated with Latrunculin A. This proved that depolymerizing the tiny actin filaments (about 37 nm length [29,30]), gave rise to an overall membrane softening. Healthy RBCs also showed heterogeneous flickering maps with multiple hot spots; however, their amplitude was significantly reduced with respect to the softened RBCs treated with LatA. By suppressing intracellular ATP concentration, the activity and metabolic energy production of RBCs treated with iodoacetamide and inosine was inhibited, resulting in a significant flickering reduction compared to healthy cells. Finally, GA fixation induced protein cross-linking by covalently binding amino groups in membrane and cytoskeletal proteins. This treatment led to irreversible rigidification of the erythrocyte membrane and structural stabilization of hemoglobin. Fixed RBCs served as a crucial control, representing a passive baseline against which active phenomena can be compared. The flickering in fixed RBCs showed minimal residual fluctuations, ascribed to passive, thermally driven fluctuations of the membrane that were locally confined by its structural integrity. Overall, non-invasive flickering mapping enabled precise differentiation of RBCs biomechanical states based on the observed spontaneous fluctuations. When the OTs were turned on, we applied on each single cell eight optical traps positioned at nearly equispaced locations on the RBCs membrane in its equatorial plane. We found that OTs operating at laser power equal to $P_1$=1.5 mW/trap induced a weak optical trapping exerting a local force lower than 1.0 pN with an experimental resolution of 0.02 pN given by the SENSOCELL detector. By comparing the flickering of free-standing and trapped RBCs, we proved that laser power $P_1$ exerted a sufficiently small optical force that did not affect the intrinsic flickering of untrapped RBCs, independently of the cell rigidity or metabolism condition. The only exception was the case of softened RBCs treated with LatA, which showed a slight reduction of flickering amplitude with respect to the highly dynamic free-standing state, as reported in Figure 4. Crucially, the observation of enhanced fluctuation amplitude at the localized trap sites, reported in Figure 3 c) and 4 for healthy, LatA-treated, and ATP-depleted RBCs in specific power states, could suggest a non-linear mechanosensing response of the active membrane-cytoskeleton complex. This effect may be indicative of a local stress-induced cytoskeletal rearrangement or active energy injection.

Our methodology confirmed that the optical traps acted as non-invasive probes, that exhibited negligible perturbation or phototoxicity. The employed 1064 nm IR laser at 1.5 mW, operated near a water absorption minimum, making it unlikely to cause significant heating in aqueous solutions [39,40]. Moreover, this minimal invasiveness was validated, as the cell's global fluctuation distribution recovered statistically after trap removal (Supplementary Figure S10), confirming that OTs induced a slight and reversible mechanical perturbation without altering the cell's intrinsic activity. Our dynamic mapping approach, using low-power OTs to measure force fluctuations and video-microscopy to retrieve membrane displacements, provided a framework for characterizing the mechanobiological states of RBCs. By reporting these two-channel measurements in the phase space of force and flickering variance, spanning multiple membrane positions for different RBCs samples, we captured the interplay between membrane dynamics and cytoskeletal integrity. This analysis revealed distinct biomechanical signatures for healthy and chemically perturbed RBCs, as reported in Figure 5. The different patterns observed in the phase-space coordinates [$\sigma_{\delta F}$, $\sigma_{\delta h}$] highlighted the sensitivity of our approach to changes in membrane rigidity and metabolic activity induced by chemical treatments. Notably, the high linear correlation (Pearson coefficient > 0.5) found between displacement and force fluctuations in healthy RBCs suggested a tightly coupled membrane-cytoskeleton system, which was altered by perturbations affecting actin structure or energy-dependent processes that showed an absence of correlation (see Supplementary Figure S12). The assessment of mechanical work produced during membrane flickering against the optical traps, as depicted in Figure 6, established a direct link between the observed mechanical dynamics and energy dissipation. Notably, LatA-treated RBCs exhibited the highest mechanical power output, exceeding even that observed in healthy cells. This result may appear counterintuitive, since latrunculin A disrupted the actin cytoskeleton and thereby reduced cortical rigidity. However, the increased power dissipation did not arise from greater active force generation; instead, it resulted from improved energy transduction caused by pronounced mechanical softening of the membrane–cytoskeleton complex. The disassembly of actin filaments possibly relieved structural constraints, that could limit fluctuation amplitude, thereby allowing the same active forces, produced by residual ATP-dependent processes, to generate larger, less-damped membrane flickering. Consequently, the higher dissipated power reflected a state of increased mechanical mobility and reduced efficiency in storing elastic energy within the spectrin–actin network. Additionally, the observed redistribution of fluctuation hot-spots and greater membrane–cytoskeleton slippage favored viscous dissipation. In summary, LatA-treated RBCs dissipated more power not because they generated more active energy, but because softening and partial uncoupling of the cytoskeleton altered the transmission pathway, enabling a larger fraction of the available energy to be irreversibly dissipated into the lipid bilayer rather than being stored in elastic energy associated with the cytoskeletal structure.

**Conclusions**

This proof-of-concept study validated the suitability of low-power optical tweezers for non-invasive force measurements, as their minimal impact on RBCs flickering avoided biomechanical artifacts, unlike higher-power settings or other probe strategies. Our findings demonstrated that RBCs membrane dynamics were strongly modulated by cytoskeletal structure and metabolic activity, revealed through correlation analysis in the force-displacement variance space. Cells treated with LatA exhibited the largest fluctuations, reflecting a loss of cytoskeletal constraint, while healthy cells showed moderate fluctuations, and ATP-depleted or fixed cells displayed minimal activity. The distinct patterns highlighted the structural role of actin in stabilizing the membrane and the energetic role of ATP in driving active fluctuations, underscoring the finely tuned balance between cytoskeletal rigidity and metabolic activity in regulating membrane mechanics and cellular adaptability. Furthermore, our analysis revealed that increased flickering in softened membranes (e.g., LatA-treated RBCs) corresponded to higher power dissipation, indicating that enhanced dynamics were energetically costly. This non-invasive methodology can enhance our understanding of the physics of biomembrane dynamics. The mechano-dynamic signature identified here represents a pre-morphological indicator of RBCs health, able to detect deviations in metabolic or cytoskeletal activity before overt shape changes could emerge. This capability lays a foundation for diagnostic applications, with the potential to identify pathological states characterized by altered membrane properties, such as hemolytic anemia. Since both local forces and membrane displacements were acquired non-invasively and analyzed algorithmically, the approach is inherently scalable. Standard OTs and video microscopy setups, coupled with automated segmentation and machine-learning classifiers, could enable high-throughput clinical screening of biomechanical phenotypes in hematological or metabolic disorders. Future studies could apply this technique to explore dynamic responses under diverse physiological conditions, quantify additional parameters such as membrane viscoelasticity, and extend its use to a wider range of pathological conditions to further elucidate the molecular mechanisms governing RBCs behavior. These prospects establish a pathway toward mechanical diagnostics at the single-cell level.

**Data and code availability**

Raw data and any additional information required to reanalyze the data reported in this paper is available from the lead contact upon request.

**MATERIALS & METHODS**

**Sample preparation.** RBCs were extracted from a single healthy donor by venipuncture employing the finger-prick method. 20 µL of blood were suspended in 500 µL phosphate saline buffer with glucose denoted as PBS+ (130 mM NaCl, 20 mM $Na_3PO_4$, 10 mM glucose, and 1 mg $mL^{-1}$ bovine serum albumin, BSA, pH 7.4) at 37 °C. The inclusion of glucose provided the necessary metabolic substrate to ensure endogenous ATP generation via glycolysis, maintaining cellular viability and physiological activity, while BSA prevents cell aggregation. The erythrocyte concentrate was obtained after 3 cycles of centrifugation of 8 min at 5000 rpm. The supernatant was discarded, and the pellet rinsed in PBS+ (500 µL). A 1:20 dilution of this suspension was maintained at 37 °C and used within a maximum period of 24 hours. Samples were mounted in a custom-built chamber (see Figure S1) assembled by using a standard microscope glass slide (75 mm x 25 mm x 1 mm-thick); then placing in the middle of the slide a square-shaped well, made of double-side adhesive tape (approximately 10 mm x 10 mm x 0.05 mm-thick), and sealing the chamber with a thin cover-glass (55 mm x 22 mm x 0.17 mm-thick, thickness code #1.5) attached to the adhesive tape. Before sealing, aliquots of 30 µL from the 1:20 suspension were poured into the well. The chamber volume filled by the RBCs suspension was therefore equal to (10 x 10 x 0.05) mm$^3$. No specific chemical treatment was applied to the glass slides or cover-glasses prior to the experiments.

We exclusively analyzed discocyte RBCs, as the presence of echinocytes was negligible. To investigate the contribution of cytoskeletal structure and metabolic activity to RBCs membrane mechanics, RBCs were subjected to three distinct biochemical treatments: ATP depletion, chemical fixation, and actin depolymerization. To inhibit ATP production and assess the role of metabolic energy in maintaining membrane dynamics, RBCs were incubated in phosphate-buffered saline lacking glucose (PBS-) supplemented with iodoacetamide (18 mM) and inosine (30 mM) for 3 hours at 37 °C. Since iodoacetamide is an alkylating agent, it irreversibly blocked glycolysis by targeting essential –SH groups in enzymes such as glyceraldehyde-3-phosphate dehydrogenase and hexokinase. Inosine acts synergistically by depleting intracellular phosphate and impairing ATP regeneration. Under these conditions, glycolytic flux was suppressed, and cellular ATP levels were drastically reduced, leading to diminished phosphorylation-dependent activity at the membrane and cytoskeleton. As a passive control, RBCs were chemically fixed using 2% glutaraldehyde (GA) for 10 minutes at room temperature. GA crosslinks membrane proteins and cytoskeletal components, immobilizing the cell and eliminating active membrane fluctuations. To disrupt the actin cytoskeleton, RBCs were treated with 1 µM Latrunculin A (LatA) for 4 hours at 37 °C in a PBS+ medium. LatA binded monomeric G-actin with high affinity, preventing its polymerization and promoting disassembly of filamentous F-actin. This treatment compromised the structural connectivity of the cytoskeletal meshwork, thereby reducing cortical tension and increasing membrane deformability and fluctuation amplitude.

All the experiments were performed at room temperature (25 °C), a compromise chosen to preserve trap stability and optical signal fidelity during long acquisitions. At physiological temperature (37 °C), RBCs displayed enhanced metabolic and cytoskeletal activity that increased spontaneous motion and thermal drift, leading to trap instability and higher photothermal stress under continuous illumination. Operating at 25 °C minimizes these sources of noise, allowing precise calibration of trap stiffness, stable alignment over 30 s recordings, and improved signal-to-noise ratios. Room temperature condition uniformly slowed RBCs dynamics without altering the mechanochemical hierarchy of different biochemical conditions (control, ATP-depleted, LatA-treated), thereby enabling reliable comparison across samples and ensuring internal consistency.

**Optical tweezers.** The optical tweezers platform (by IMPETUX, Spain) employed acousto optic deflectors (AOD) for beam-steering deflection and a single-beam optical tweezers system with a laser emitting at $\lambda = 1064$ nm, allowing rapid multiplexing of local traps positioned on the RBCs equatorial membrane. It operated through the *LightAce* software interface, developed using LabView. This interface enabled precise activation of traps at designated locations by selecting them with the PC cursor on the live camera feed. The system was equipped with a direct force measurement device capable of detecting the change in optical momentum from the optical traps (SENSOCELL). The detector measured the deflection of the laser beam due to the interaction with the sample, which was directly proportional to the momentum exchanged, and therefore to the net force exerted. The direct force signal was continuously recorded at 25 kHz (sampling rate of the photodiode detector). At each sampling interval (0.04 ms), the AODswitched the beam to the next trapping site, where the local force was measured. Consequently, all eight trapping sites were revisited every 0.32 ms, corresponding to an effective local sampling rate of 3125 Hz per site. Thus, each recorded force time series corresponded to the local membrane interaction at specific trapping position, preserving locality during multiplexed operation. The OTs were mounted on an inverted microscope (Eclipse Ti, Nikon, Japan), using a water immersion objective (CFI Plan Apochromat VC 60X C WI; numerical aperture NA=1.2, by Nikon). This objective served a dual role: it was used for high-resolution imaging of the cell membrane and to generate the optical traps. The sample chamber was placed with the thin coverslip forming the bottom of the chamber and in direct contact with the 60X water immersion objective (see Figure S1). Conversely, the thick microscope glass slide formed the top of the chamber and was in contact with the oil-immersion high numerical aperture lens (NA=1.4, by IMPETUX). This lens acted as a condenser, which was simultaneously used for providing white light illumination and for collecting the transmitted laser light to measure the change in optical momentum by the force detector.

**Microscopy setup.** The sample chamber was mounted on a two-axis translation stage for precise alignment at the sub-micron scale. The collection objective was the same used for inducing the optical tweezers. The bright-field images were recorded by the CMOS camera (ORCA-spark, Hamamatsu, Japan) at 65 Hz frame rate (integration time ~15 ms) and with duration of 2000 frames for each video, corresponding to ~30 s. The camera acquisition was hardware-triggered by the SENSOCELL force detector, ensuring time alignment between imaging and force recording.

**RBCs trapping protocol.** After approximately 30-minute sedimentation period, sufficient for gravity to deposit the cells, the RBCs achieved weak adhesion to the glass chamber bottom coverslip through unspecific interactions. The small adhesion zone occurred solely at the bottom basal surface of the discocyte, providing sufficient immobilization for stiffness characterization, while leaving the membrane in the equatorial plane free to fluctuate. In free-standing conditions (i.e., with the optical trap deactivated), we recorded a video capturing the RBCs flickering motions under healthy physiological conditions. During the flickering measurements, 8 optical traps were manually positioned along the RBC membrane contour in its equatorial plane using the SENSOCELL software. Due to the inherent morphological variability of each cell and the manual nature of the positioning process, the angular separation between the traps was not strictly uniform. Instead, the traps were strategically distributed across the quadrants of the cell membrane, ensuring a distributed force application in geometrically similar regions from cell to cell. All RBCs were recorded firstly free-standing, then optically trapped at lower and higher laser power. During the experiment, they were consistently positioned near the center of the microscope observation field. This practice minimized any residual positional differences between force detection relative to the microscope's reference system. In addition, visual inspection of the samples under the microscope during the entire experimental session (about 3–4 hours) did not reveal any noticeable increase in echinocytes formation or other morphological alterations of the RBCs (estimated to be approximately less than 5 % of the total cells).

**Flickering analysis.** Each frame of the video acquired by the fast CMOS camera was saved as TIF file and analyzed by WOLFRAM MATHEMATICA software to evaluate the position of the circular RBC membrane as a function of time. The rim halos at the RBCs membrane allowed us to estimate the radial positions of the contours by interpolation at subpixel resolution with an accuracy of few nm, see Supplementary Figure S3. We recorded the time series of the local membrane fluctuations for 2048 segmented elements, corresponding to an angular step $\Delta\theta \approx 3.1 \cdot 10^{-3}$ rad, and to an arc length of 12 nm. The separation between adjacent points detected on the external membrane, was smaller than the estimated size of the flickering unit, defined as the smallest statistically independent spatial segment of the membrane contour (approximately 50-100 nm) that contributed a localized fluctuation signal. By tracking the membrane position over time, we evaluated the fluctuations $\delta h(t, \theta)$ and corrected global drift that the RBC might experience due to global translation or rotation. To remove global translations, we evaluated the in-plane position of the center of mass of each tracked contour and subtracted its trajectory from the contour coordinates. Subsequently, and only after the removal of translations, we eliminated rigid rotations by evaluating the best alignment between consecutive frames, i.e., the rigid transformation that conserved the distances among the traced membrane elements.

**Data processing.** The effective stiffness constant of the trap was estimated by using a particle scan routine included in the *LightAce* software, that controlled the trap displacement while measuring the local force exerted on the RBCs membrane. Prior to all measurements, the system's power calibration routine was performed to actively correct for the non-uniform transmittance of the optical system and ensure laser power stability across the entire trapping region (70 μm x 70 μm), achieving a measured error of less than 2%. The SENSOCELL detector measured directly the $x, y$ in-plane components of the trapping force at a given trap position $^lF_x^{trap}$ and $^lF_y^{trap}$. By removing the background contribution, we evaluated the in-plane components of the force exerted by the membrane at position $l$: $^lF_x = -\left(^lF_x^{trap} - ^lF_x^{off}\right)$ and $^lF_y = -\left(^lF_y^{trap} - ^lF_y^{off}\right)$. To obtain the relevant force component that aligns with the local radial membrane fluctuations ($\delta h$) measured by video microscopy, we transformed the cartesian components into radial ($^lF_R$) and tangential ($^lF_T$) components with respect to the membrane contour by determining the local angle ($\theta$) at each trap position, as highlighted in Figure S3. Therefore, by applying the rotation we obtained: ${}^lF_R = {}^lF_x\cos\theta + {}^lF_y\sin\theta$ and ${}^lF_T = -{}^lF_x\sin\theta + {}^lF_y\cos\theta$. We found that $|{}^lF_R| \gg |{}^lF_T|$ at every trap position and for every studied RBC. Since ${}^lF_R$ is the force component parallel to $\delta h$, we used only the radial force component for calculating the force fluctuations $\sigma_{\delta F}$ and the work produced ${}^lW(t)$. Force and displacement time series were triggered via a shared hardware and were acquired in aligned timestamps. Temporal data were recorded using custom LabVIEW routines within the same environment. For each trapping site, the time series of the local radial force fluctuations ${}^l\delta F_R(t)$ was extracted only from the intervals when the beam was at that specific site, preserving the local sampling rate of 3125 Hz. The membrane contour was independently tracked from bright-field videos acquired at 65 Hz; local displacement fluctuations, ${}^l\delta h(t)$, were computed for eight trap positions in each cell. Regarding the data reported in Figure 5, both $\sigma_{\delta F}$, $\sigma_{\delta h}$ were calculated over the same, co-registered time windows, aligning both modalities without altering the force bandwidth, i.e., no temporal averaging of the faster force signal was applied. For each sample and RBC, the ($\sigma_{\delta F}$, $\sigma_{\delta h}$) pairs from the eight trap positions were aggregated to construct global dynamic maps. For the analysis reported in Figure 6 a), the high-frequency radial force data was integrated over the video acquisition interval to match the timescale of the displacement data. Specifically, the force values were averaged during the image integration time, resulting in a time series with the same temporal resolution of ${}^l\delta h(t)$. This temporal averaging step was crucial for the Stratonovich discretization used in the work calculation.

**Thermodynamic analysis.** Stochastic work done by the membrane against the trap was computed for each trap position along the radial direction using the Stratonovich (mid-point) convention [36]. Let $\{t_m\}_{m=0}^n$ be the time grid defined by video-microscopy acquisition at 65 Hz, with $\Delta t_m = t_{m+1} - t_m$. The local variation of the membrane radius for a given trap position is $\delta R_m = R(t_{m+1}) - R(t_m)$. For each interval $[t_m, t_{m+1}]$, we evaluated the mid-point force $\bar{F}_R^{(m+1/2)}$ by averaging within the interval the local force samples recorded at 3125 Hz:

$$\bar{F}_R^{(m+1/2)} = \frac{1}{\Delta t_m}\int_{t_m}^{t_{m+1}} F_R(t)\, dt \approx \frac{1}{M_m}\sum_{k=1}^{M_m} F_R(t_{m,k}),$$

where $t_{m,k}$ are the timestamps at which force was recorded in the interval $[t_m, t_{m+1}]$, and $M_m$ is the number of force data (3125/65 ~ 48) collected during $\Delta t_m$, when the beam was active at site $l$. The elementary work, $\delta W_m$, performed in the time interval $[t_m, t_{m+1}]$, and the cumulative work, $W$, can be obtained as:

$$\delta W_m = \bar{F}_R^{(m+1/2)}\, \delta R_m \quad ; \quad W = \sum_{m=0}^{n-1} \delta W_m$$

which is the generalization for discrete Stratonovich approximation:

$$W(t) = \int_0^t F_R(t') \circ dR(t') \cong \sum_{m=0}^{n-1}\left(\frac{F_R(t_m) + F_R(t_{m+1})}{2}\delta R_m\right)$$

The mean dissipated power over a window of duration $\Delta t$ was given by $P \equiv W(\Delta t)/\Delta t$.
As a robustness check, using an Itô (left-point) Riemann sum or a central-difference variant yielded work and power statistically indistinguishable within experimental uncertainty, confirming that our results were dominated by low-frequency membrane dynamics and were insensitive to the precise discretization.

**Statistical Analysis.**
For the healthy, LatA and ATP- samples we analyzed N=10 RBCs, while for the fixed sample only N=7 RBCs were studied. The normalized probability density functions (PDFs) of the local fluctuation, $\delta h$, were calculated for all samples as the total fluctuation events distribution normalized to their integral. Specifically, the PDFs were constructed by considering every single fluctuation event measured over the 30 s observation time at 65 Hz, for all 2048 tracked membrane locations and all N cells per sample. This procedure resulted in approximately $10^7$ total data per PDF. To generate the distributions of the flickering amplitude, from each cell 2048 positions were tracked as a function of time, giving rise to 2048 values of $\sigma_{\delta h}$, yielding a total of $\sim 2\times10^4$ data per sample that were used to construct the reported distributions. In contrast, for the distributions evaluated only at trap positions (orange data in Figures 3 and 4), we extracted only 8 data points per cell, i.e., one for each optical trap. Therefore, the distributions were made of 8xN ~ 80 data for each sample. The statistical significance of differences between distributions of different samples were evaluated by a pairwise two-sample $t$-test at the level of significance set to or 0.01, as specified in the main article text. Software used for statistical analysis: MATLAB and OriginPro 2023.

**Acknowledgments:** The authors acknowledge Grants CNS2023-143803, PID2019-108391RB-100 and TED2021-132296B-C52 funded by MICIU/AEI/10.13039/501100011033 and by "European Union NextGenerationEU/PRTR". The authors acknowledge Felix Ritort and Frederic Català i Castro for fruitful discussions and technical support. We thank the anonymous reviewers for their insightful and constructive comments, which substantially improved the clarity and rigor of the manuscript.

**Author Contributions:** A. D. prepared the samples, conducted the experiments, and analyzed the data, supported by C.L. and N.C. D.H.A. designed computational tools. M.C. performed controls of four samples. N.C. provided computational tools, interpreted the results, drafted the manuscript and figures. N.C. and F.M. designed the research, conceived the experiments, interpreted the results, and prepared the final manuscript. F.M. provided equipment and infrastructure. All authors contributed to the manuscript.

**Declaration of interests:** The authors declare no competing interest.

**Declaration of generative AI and AI-assisted technologies in the writing process:** During the preparation of this work the authors used ChatGPT 3.5 and Gemini 2.5 Pro to check English grammar and increase the readability of the text in some parts of the paper. After using this tool, the authors reviewed and edited the content as needed and took full responsibility for the content of the publication.

# Supplementary Information for

# Active Force Dynamics in Red Blood Cells
# Under Non-Invasive Optical Tweezers

Arnau Dorn[1,2], Clara Luque-Rioja[1,2], Macarena Calero[2], Diego Herráez-Aguilar[3], Francisco Monroy[1,2,#] and Niccolò Caselli[1,2,*]

[1] *Departamento de Química Física, Universidad Complutense de Madrid, Ciudad Universitaria s/n, 28040 Madrid, Spain*
[2] *Translational Biophysics, Instituto de Investigación Sanitaria Hospital Doce de Octubre, 28041 Madrid, Spain*
[3] *Instituto de Investigaciones Biosanitarias, Universidad Francisco de Vitoria, Ctra. Pozuelo-Majadahonda, Pozuelo de Alarcón, Madrid, Spain*

**Email:** *ncaselli@ucm.es; #monroy@ucm.es

**This PDF file includes:**

Supporting Information Figures S1 to S13

Supporting Information Tables T1 to T10

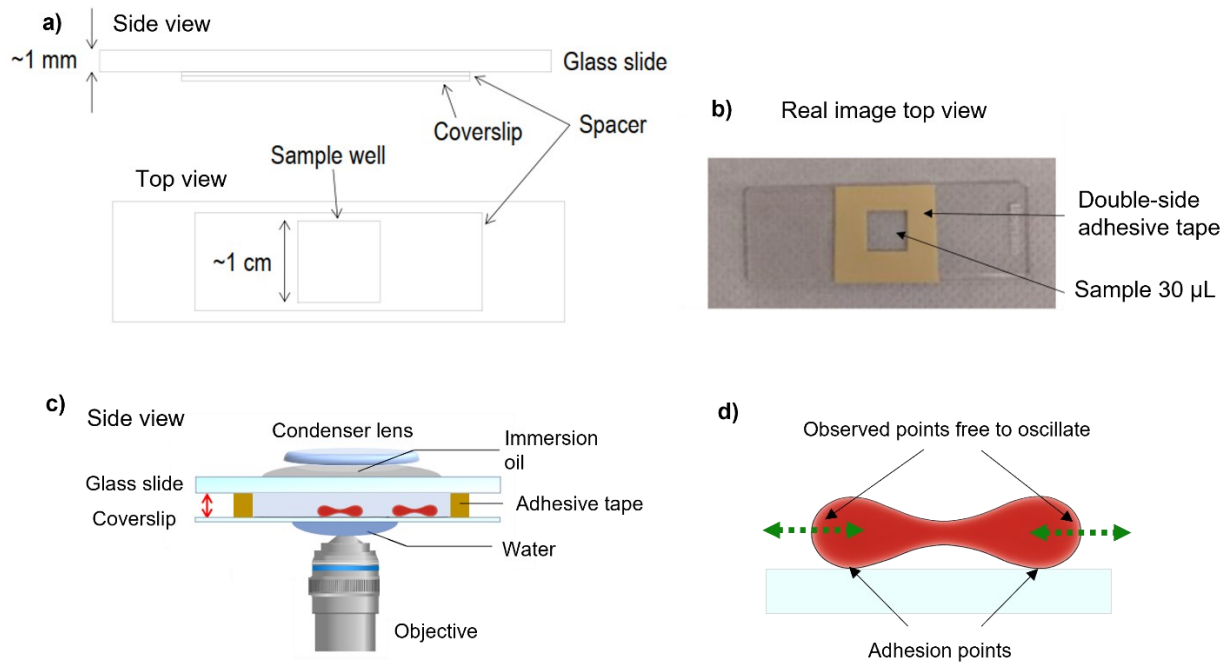

**Figure S1. Schematics of the sample chamber and mounting configuration. a)** Side and top view schematics of the sample chamber used in the experiments. **b)** Real image of the sample chamber (top view). **c)** Schematic side view (not in scale) of the experimental setup: the RBCs dilution (with discocyte shaped cells) was placed in the sample chamber under the microscope. **d)** Schematic side view of a single RBC resting on the bottom coverslip glass. Adhesion occurred locally at the bottom basal surface of the discocyte due to gravity and unspecific surface interactions. The membrane fluctuations measurements were performed in the equatorial plane of the cell, where the membrane was free to oscillate (green arrows). These positions were ~µm distant from the adhesion points both in the equatorial plane and vertical direction. This configuration is referred to as free-standing in the main manuscript because optical traps were not applied.

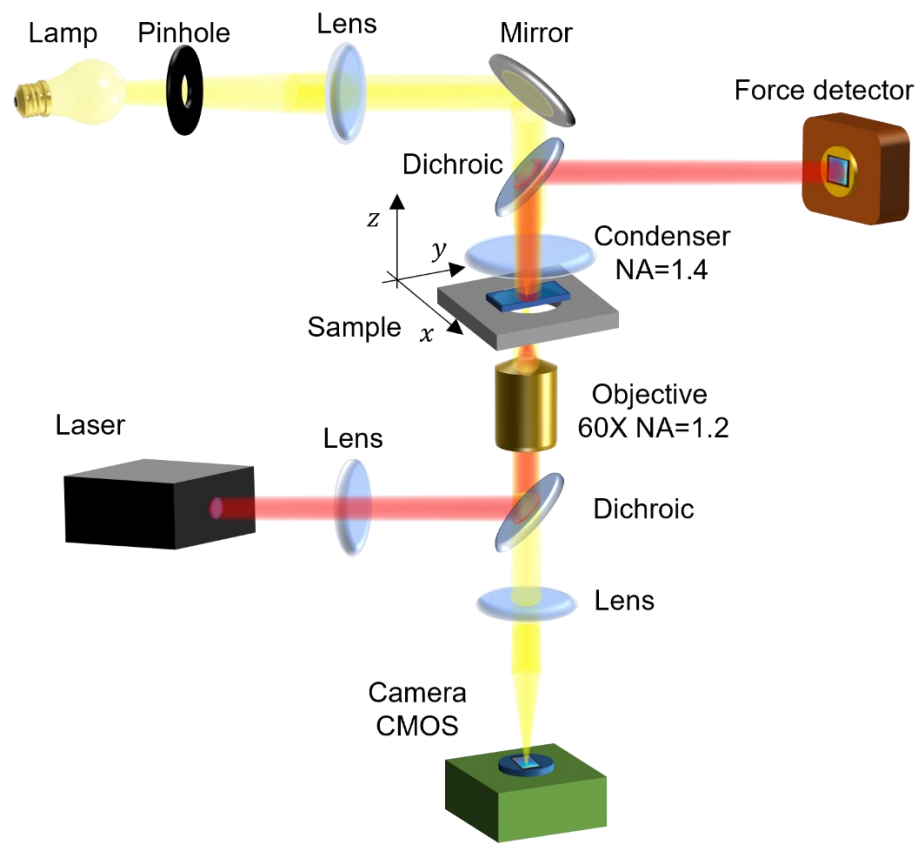

**Figure S2. Schematics of the experimental setup.** The optical path of the optical tweezers is reported as the red beam, and the optical path of bright field video-microscopy is reported as the yellow beam.

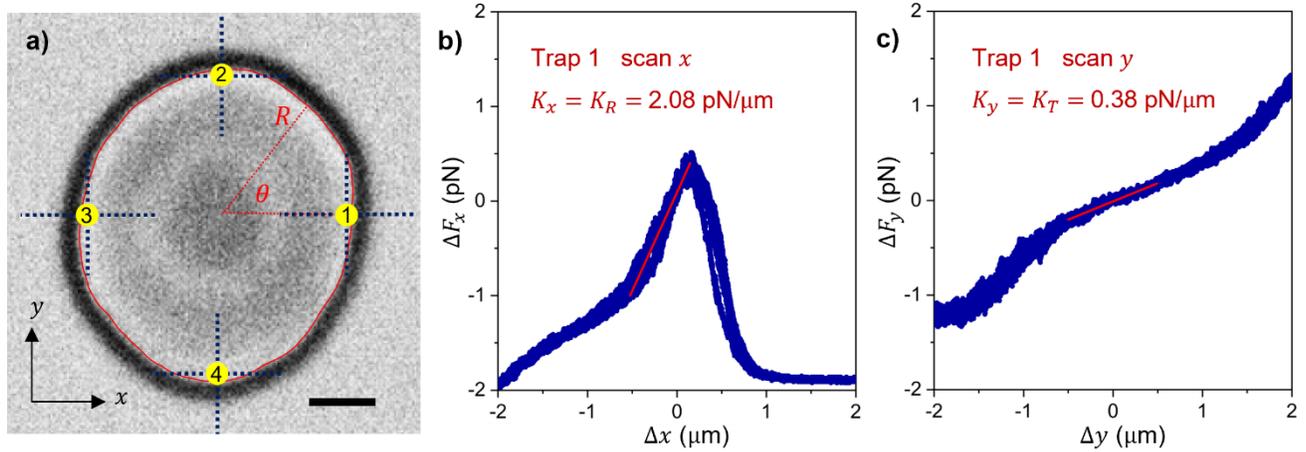

**Figure S3. Stiffness of optical tweezers directly applied to the RBC membrane. a)** Bright-field image of a healthy RBC observed in the equatorial plane $xy$. The red curve represents the membrane contour evaluated by the membrane tracking algorithm as the rim halo. $R$ and $\theta$ are the local cell radius and angle with respect to the center of the cell. They were employed to evaluate the radial displacement $\delta h(\theta, t) = R(\theta, t) - \langle R(\theta, t) \rangle$ and flickering amplitude $\sigma_{\delta h}$ along the membrane contour. The Scale bar is 2 μm. The four yellow dots distributed along the RBC membrane rim are the positions where single optical tweezers were applied by focusing a Gaussian laser beam with power P$_1$=1.5 mW/trap at successive times. To retrieve the local trap stiffness, we induced a controlled displacement to each single trap by moving the laser beam. This task was accomplished by means of the SENSOCELL integrated software, which employed acousto-optic deflectors to execute the scan in few milliseconds. Dashed lines represent the trap displacements of ±2 μm in $x$ and $y$ directions. During this scanning procedure, we visually verified that the whole cell did not move globally and adhered to the glass surface, confirming that the measured force response was due to the membrane's local resistance and the optical trap itself. **b)-c)** Optical trapping force measured when the OTs were applied at position 1, as a function of the trap displacement $\Delta x$ and $\Delta y$, respectively. The values of $\Delta x = 0$, $\Delta y = 0$ correspond to the intersection of the dashed lines at location 1. For this specific location, the force $\Delta F_x$ is the radial component while $\Delta F_y$ is the tangential one. Each scan was performed unidirectionally 5 times, starting from the minimum displacement of -2 μm, to ensure reproducibility and all the data from those measurements were reported as scatter dots in **a)** and **b)**. The force values were reported by subtracting the initial offset force measured at the trap's starting position (by defining $\Delta F_x = 0$ at $\Delta x = 0$ and $\Delta F_y = 0$ at $\Delta y = 0$). Red lines are the linear fit of the force data in the proximity of the initial trap position, in the range $(-0.5, +0.5)$ μm. Trap stiffness was calculated as the fitting line slope in this range, assuming a harmonic trapping potential ($\Delta F_i = -K_i \Delta x_i$). Therefore, for the trap in position 1, the stiffness calculated along $x$ corresponds to the radial contribution of the RBC membrane ($K_x = K_R$) and along $y$ to the tangential component ($K_y = K_T$). The values of radial and tangential stiffness obtained in **b)-c)** were $K_R = 2.08$ pN/μm and $K_T = 0.38$ pN/μm, respectively. The same evaluation was performed for traps placed in positions 2, 3 and 4. The radial stiffness was consistently larger than the tangential stiffness at all positions. This difference can be attributed to the fact that radial trap displacement experienced a much larger variation in the refractive index spatial distribution (when crossing the boundary between intracellular and external medium) with respect to the tangential scan, in which the laser did not reach the external medium. The different spatial variation of the refractive index resulted in a greater contribution of the optical gradient force to $K_R$, with respect to $K_T$.

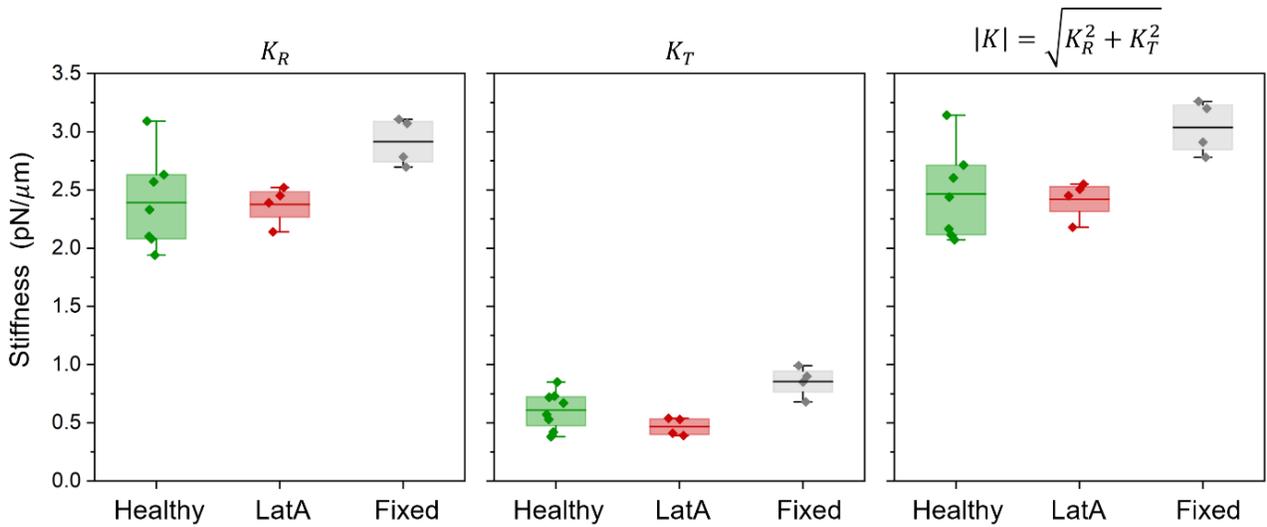

**Figure S4**. **Optical tweezers stiffness**. For each single optical trap, the stiffness was evaluated using the method described in Figure S3, employing a laser power equal to $P_1$=1.5 mW/trap. $K_T$ represents the stiffness obtained for scans tangential to the RBC membrane. $K_R$ is the stiffness obtained for radial scans, normal to the RBC membrane contour. The stiffness modulus was defined as $|K| = \sqrt{K_R^2 + K_T^2}$. The reported measurements were obtained for three RBCs conditions: healthy, LatA and fixed cells. Data corresponding to healthy RBCs were extracted from 4 measurements on 2 different cells (8 total data). Data for LatA-treated and Fixed RBCs were generated from 4 measurements on 1 single cell. The primary objective of these measurements was to establish the correct order of magnitude of the stiffnesses used for force detection, as the main interest in this work focuses on force fluctuations.

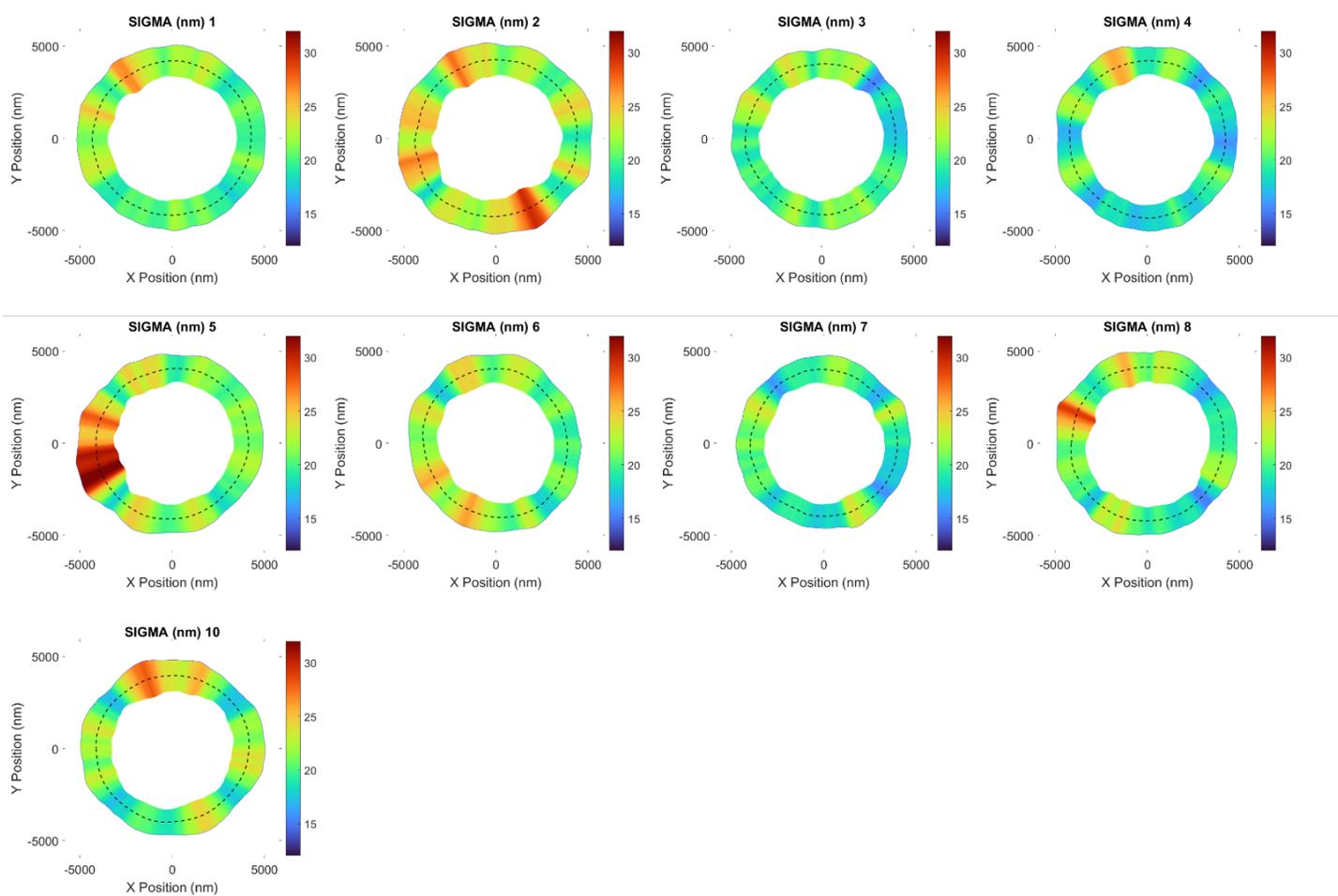

**Figure S5. Flickering maps for free-standing healthy RBCs**. In these measurements no optical tweezers were applied. The maps were evaluated as the standard deviation of the radial membrane fluctuation, $\sigma_{\delta h}$, and reported in the same range (12-32) nm. Dashed lines represent membrane mean positions. The XY plane coordinates have their origin coincident to the center of the RBCs. The map spatial amplitude was exaggerated and out of scale. The explanation of individual quantities and symbols here defined also applies to the following Figures S6 – S8.

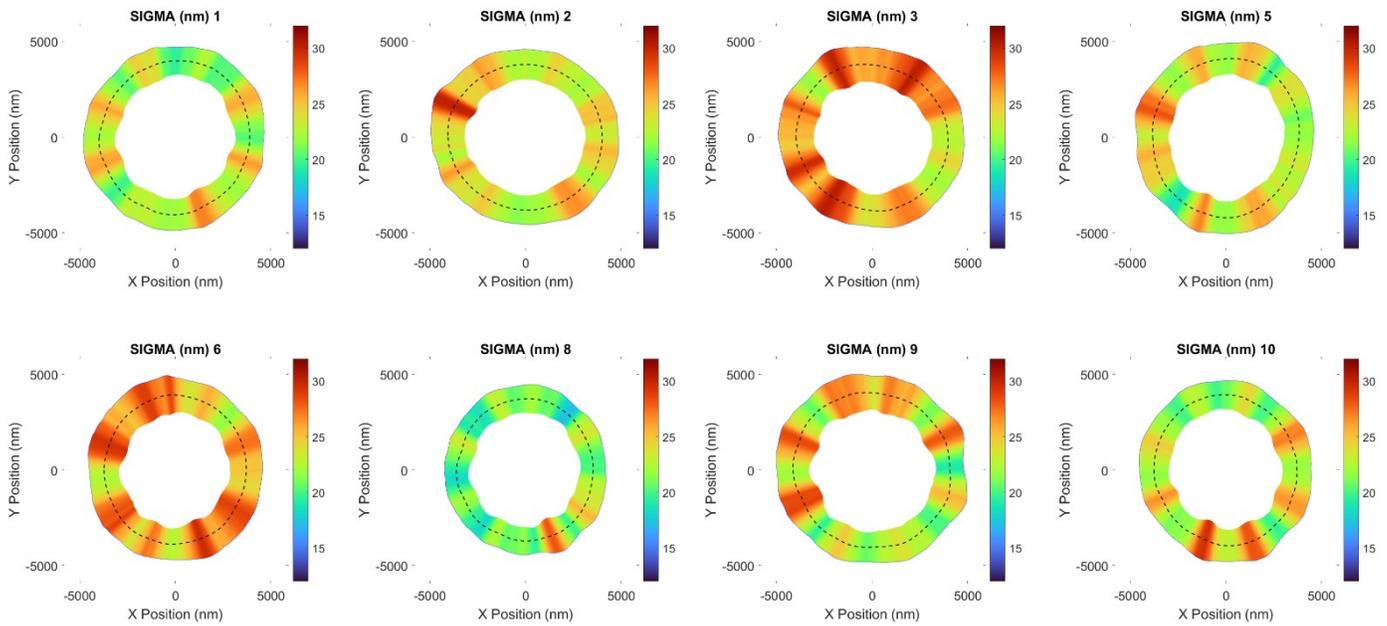

**Figure S6**. **Flickering maps for free-standing RBCs treated with Latrunculin A (LatA)**. This treatment was performed to reduce the membrane rigidity. No optical tweezers were applied during these measurements.

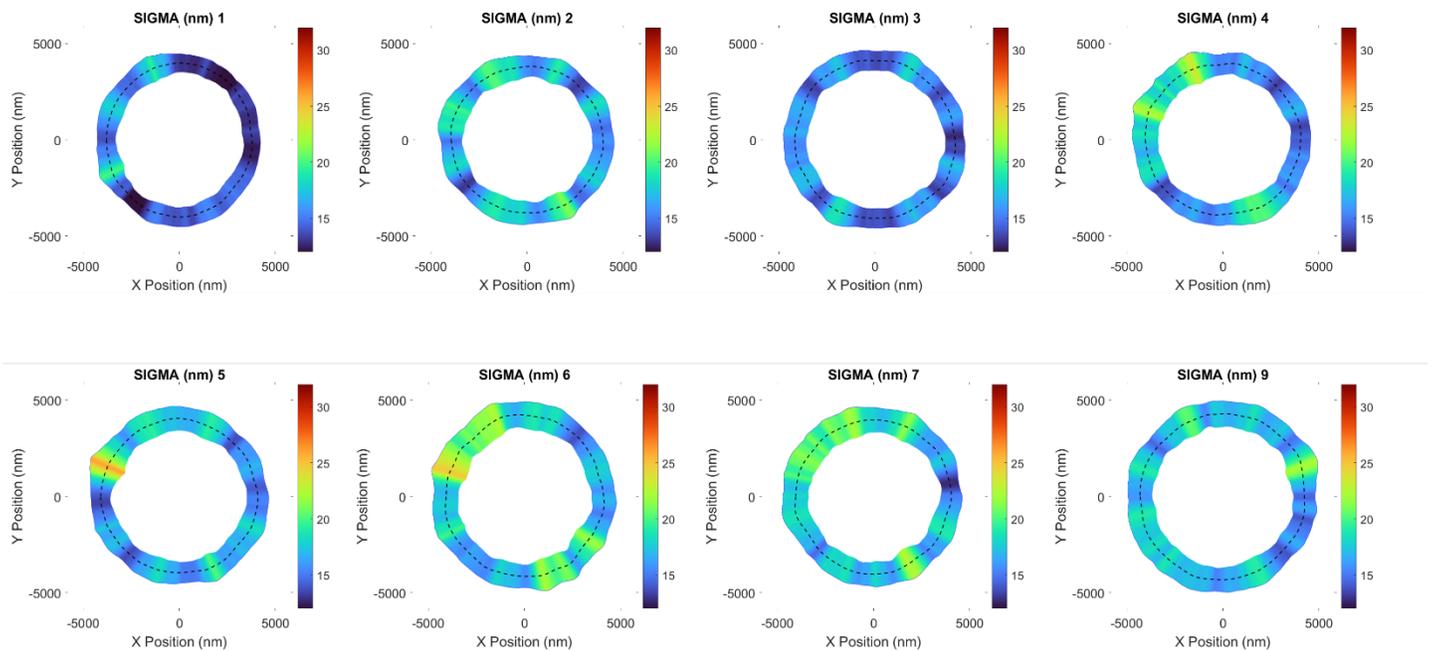

**Figure S7**. **Flickering maps for free-standing RBCs treated with inosine and iodoacetamide (ATP-)**. This treatment was performed to inhibit cytoskeletal phosphorylation and production of ATP. No optical tweezers were applied during these measurements.

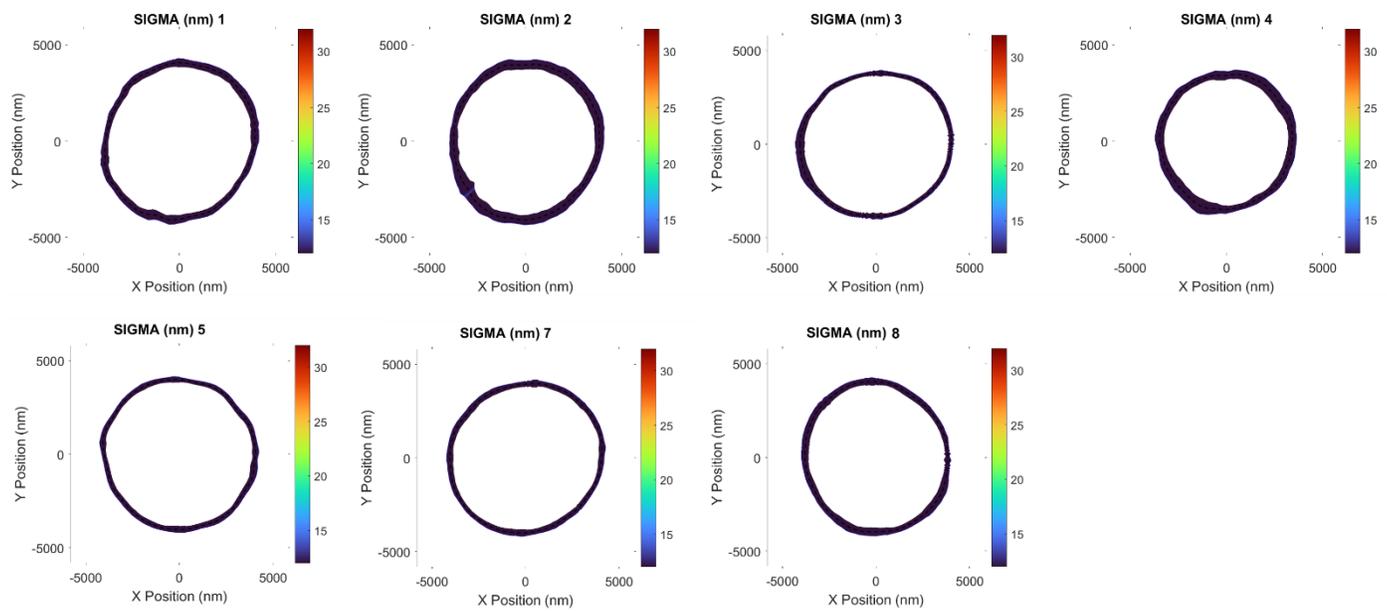

**Figure S8**. **Flickering maps for free-standing RBCs treated with glutaraldehyde (fixed)**. This treatment was performed to solidify hemoglobin and crosslink the cytoskeleton. No optical tweezers were applied during these measurements. 7 RBCs were analyzed.

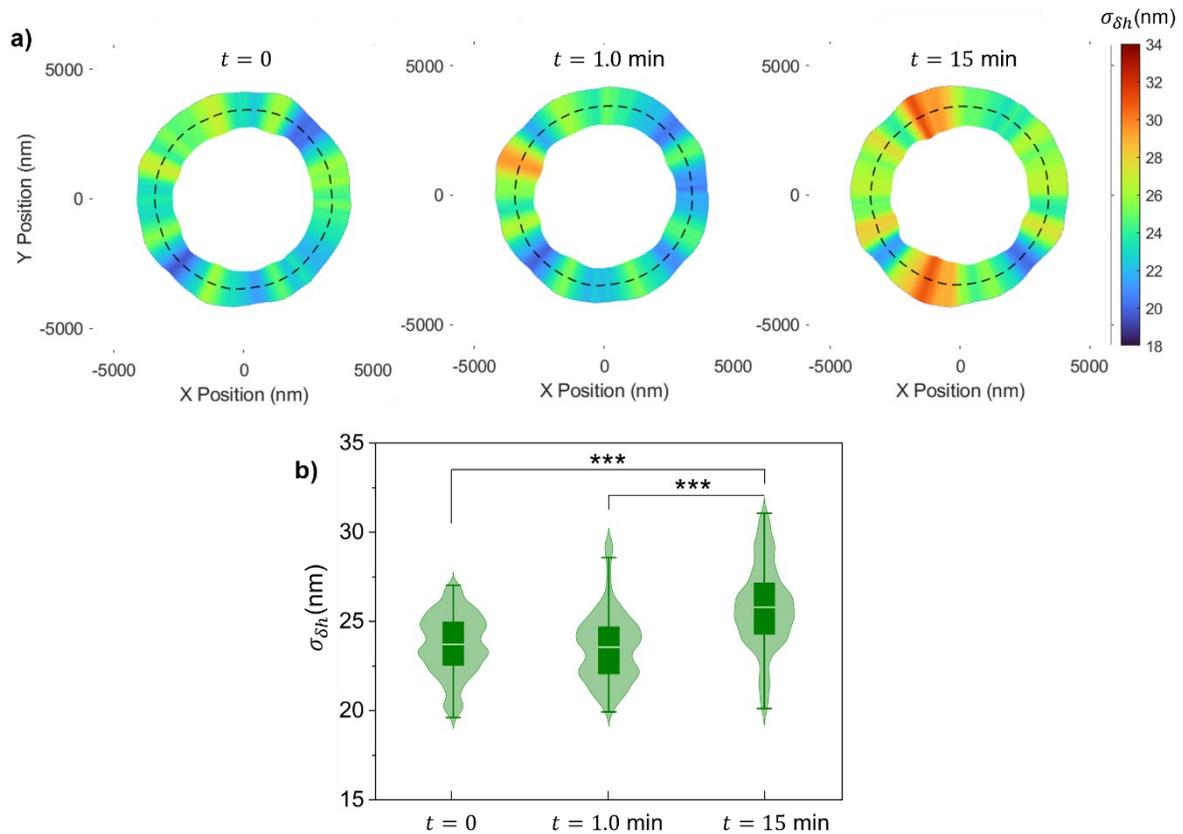

**Figure S9. Temporal reproducibility of membrane flickering and local hot spots. a)** Flickering maps, $\sigma_{\delta h}$, for a single healthy and free-standing RBC performed in time succession. The first map (left panel) was acquired at time $t = 0$, the second at time $t = 1.0$ min (central panel) and the third at time $t = 15$ min (right panel). Each measurement lasted for 30 s. **b)** Distributions of $\sigma_{\delta h}$ for the data reported in a). Statistical significance of differences between distributions was found only for the measurement at the largest time point ($t = 15$ min). The symbol *** indicates a statistically significant difference between the two labeled samples (pairwise two-sample $t$-test, $p \ll 0.001$). The pairwise comparison shows a negligible difference between the initial and the $t = 1.0$ min measurement regarding both flickering distributions and hot spots locations, thus confirming short-term persistence of flickering. Conversely, the distribution measured at time $t = 15$ min shows a significant variation indicating long-term reorganization of the cytoskeletal activity.

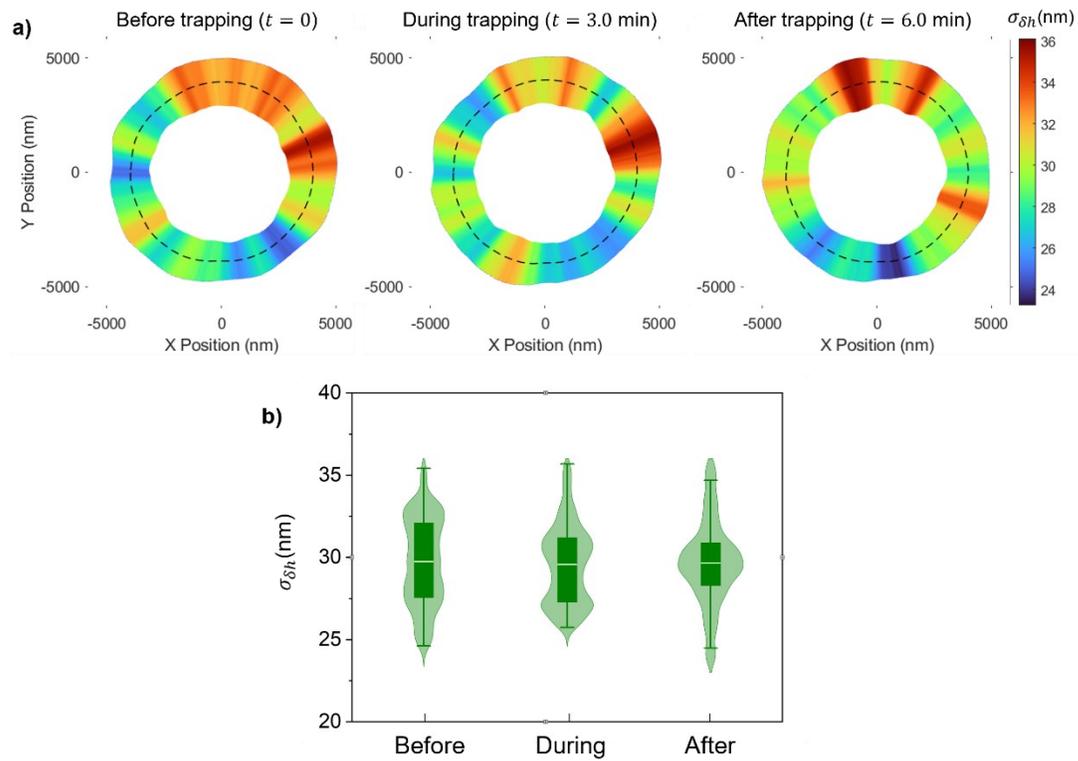

**Figure S10. Reversibility of membrane flickering after optical trapping. a)** Flickering maps, $\sigma_{\delta h}$, for a given healthy RBC performed in time succession, reported from left to right: before the application of optical trapping (free-standing cell at $t = 0$, left panel), during the application of OTs at laser power P$_1$ = 1.5 mW/trap (at $t = 3.0$ min, central panel) and after the OTs were turned off (free-standing cell at $t = 6.0$ min, right panel). Each measurement lasted for 30 s. **b)** Distributions of $\sigma_{\delta h}$ for the data reported in **a)**. No statistically significant difference was found between distributions (pairwise two-sample $t$-test showed $p > 0.05$), confirming that the mechanical effects induced by the optical traps were reversible and possibly did not damage the cell structure. However, after 6 min the local hot spots were redistributed along the membrane contour. The RBC presented in this Figure originated from a different healthy donor than the one used for the cells studied in the main manuscript.

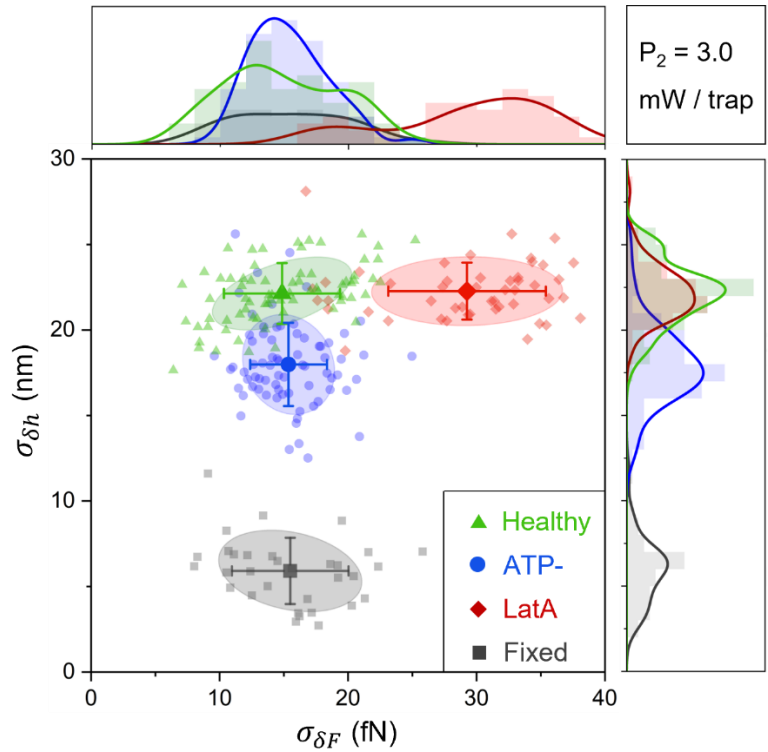

**Figure S11. Dynamics map for RBCs probed with optical tweezers at laser power P₂=3.0 mW/trap.** The vertical axis reports the standard deviation of the local membrane radial fluctuation, $\sigma_{\delta h}$, and the horizonal axis the standard deviation of the radial component of the force variation, $\sigma_{\delta F}$, measured by optical tweezers applied at the same membrane position. Different groups correspond to healthy RBCs (green triangles); RBCs treated with Latrunculin A (LatA, red diamonds); RBCs treated with inosine and iodoacetamide (ATP-, blue circles); RBCs treated with Glutaraldehyde (Fixed, gray squares). The bigger symbol in each ensemble is the mean value, error bars are the standard deviation of the group distribution. Ellipses correspond to the 70% confidence regions. The top and right distributions are the projections of the dynamics map on the $\sigma_{\delta F}$ and $\sigma_{\delta h}$ axis, respectively. Data are reported for OTs at laser power equal to P₂=3.0 mW/trap and were extracted form $N$ =10 RBCs for the healthy, LatA, and ATP- samples, and $N$ =7 RBCs for the fixed sample.

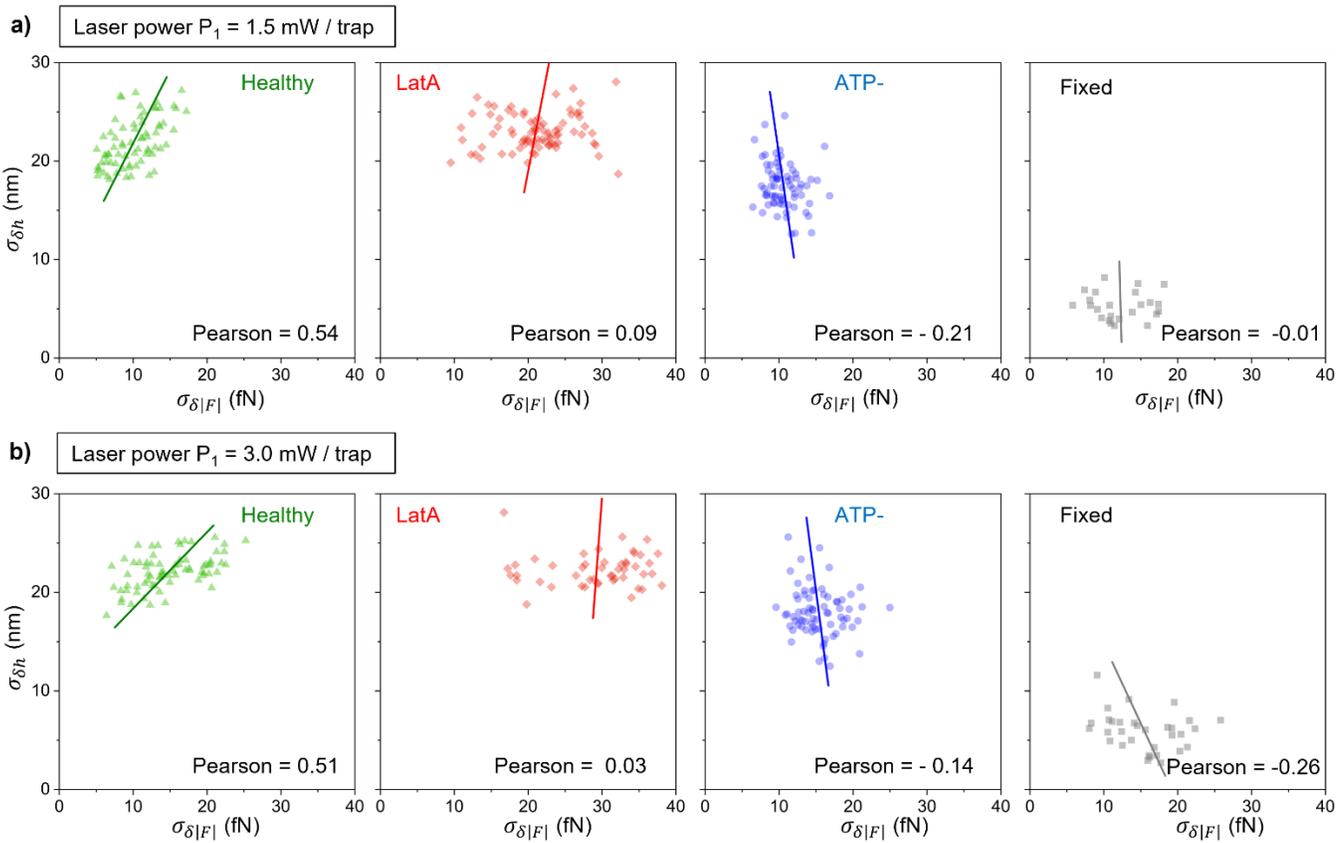

**Figure S12. Dynamics maps correlation analysis.** Data reported in Figure 5 and S11 were reported in single panels separating them for sample and laser power. The $\sigma_{\delta h}$ - $\sigma_{\delta F}$ space was reported for RBCs probed by optical tweezers at laser power $P_1$=1.5 mW/trap **(a)**, and $P_2$=3.0 mW/trap **(b)**. The Pearson linear correlation coefficient was reported for each case and the liner fit of the data was represented as a straight line in each panel.

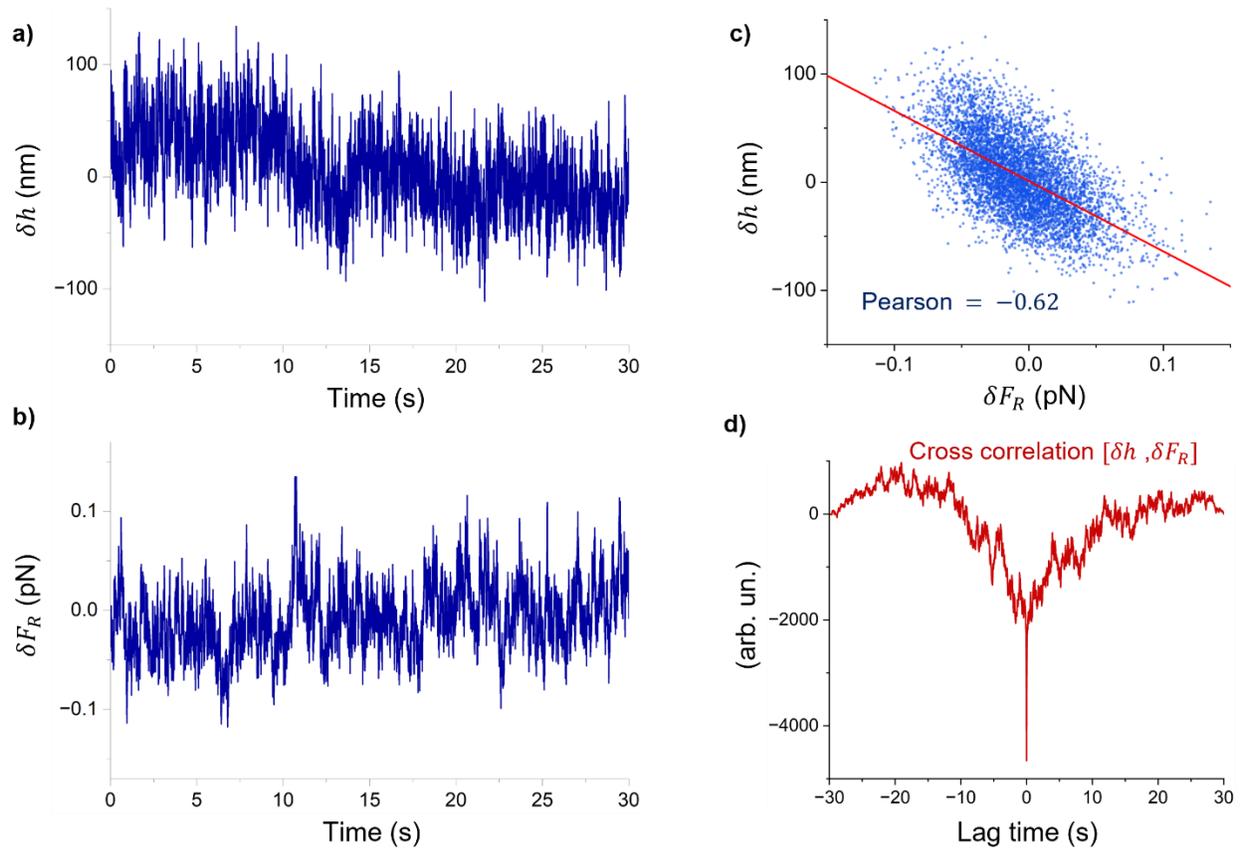

**Figure S13. Correlation between force and spatial fluctuation at a given membrane position. a)-b)** Time series of spatial fluctuations, $\delta h(t)$, and radial force fluctuations, $\delta F_R(t)$, respectively. Both signals were detected at a given trap position on the membrane of a healthy RBC at laser power $P_1$=1.5 mW/trap. Both time series were plotted with a frame rate of 65 Hz, by averaging the force data, as discussed in Methods. **c)** Data presented in a) and b) were represented in the $[\delta h, \delta F_R]$ space. Echa data point corresponds to a given acquisition time. The red line is the linear fit of the data and represent a guide to the eye for highlighting the distribution trend. The Pearson linear correlation coefficient was evaluated as equal to $-0.62$, indicating a high level of anti-correlation between the two independently measured variables, in agreement with the harmonic approximation $\delta F_R = -K\delta h$. **d)** Cross correlation between the two time-series $\delta h(t)$, $\delta F_R(t)$ evaluated as a function of the lag time by using the internal correlation function in the software *OrginPro 2023*. The symmetric cross correlation exhibited a negative minimum at zero lag time, indicating a high level of anticorrelation between the two variables. The same results indicating an anticorrelation were found for different trap positions and RBCs.

**Table T1. Stiffness Values of $K_R$, $K_T$ and $|K|$ for each RBCs sample.** All stiffnesses are visualized in Figure S4 were reported as the mean ± standard deviation. No significant differences were observed in any component between healthy and LatA-treated cells, indicating the treatment did not affect overall membrane stiffness. In contrast, fixed cells showed significantly higher stiffness values, approximately a 20% increase in $|K|$, possibly due to induced protein crosslinking. Despite these stiffness differences, the biconcave shape of the cells remained constant throughout the experiments, with the appearance of less than 10% echinocytes in every sample by the end of the measurement session (approximately 3-4 hours after sample preparation).

|  | $K_R$ (pN/μm) | $K_T$ (pN/μm) | $|K|$ (pN/μm) |
|---|---|---|---|
| Healthy RBCs | 2.4 ± 0.4 | 0.61 ± 0.16 | 2.5 ± 0.4 |
| LatA RBCs | 2.38 ± 0.18 | 0.47 ± 0.08 | 2.4 ± 0.2 |
| Fixed RBCs | 2.9 ± 0.2 | 0.86 ± 0.16 | 3.0 ± 0.2 |

**Table T2**. Analysis of the data reported in Figure 2 c) of the main manuscript. Second column: average value of $\sigma_{\delta h}$, with uncertainty given by its standard deviation, evaluated for each sample. Right box: probability ($p$-value) calculated from all the pairwise $t$-tests.

|  | $\sigma_{\delta h}$ (nm) |  | $p$-value | | | |
|---|---|---|---|---|---|---|
|  |  |  | Healthy | LatA | ATP- | Fixed |
| Healthy | 21.0 ± 2.5 |  |  |  |  |  |
| LatA | 23.8 ± 2.5 |  | $10^{-63}$ |  |  |  |
| ATP- | 16.7 ± 2.2 |  | $< 10^{-100}$ | $< 10^{-100}$ |  |  |
| Fixed | 6.8 ± 1.5 |  | $< 10^{-100}$ | $< 10^{-100}$ | $< 10^{-100}$ |  |

**Table T3**. Analysis of the data reported in Figure 3 c) of the main manuscript for healthy RBCs considering all tracked membrane positions (green data). Second column: average value of $\sigma_{\delta h}$, with uncertainty given by its standard deviation, evaluated for each sample at different trapping laser power. Right box: probability ($p$-value) calculated from every pairwise $t$-test.

|       | $\sigma_{\delta h}$ (nm) | p-value |  |  |
|-------|--------------------------|---------|---------|---------|
|       |                          | $P_0$   | $P_1$   | $P_2$   |
| $P_0$ | 21.0 ± 2.5               |         |         |         |
| $P_1$ | 21.2 ± 2.4               | 0.41    |         |         |
| $P_2$ | 21.5 ± 2.2               | $3 \times 10^{-3}$ | $7 \times 10^{-3}$ |         |

**Table T4**. Analysis of the data reported in Figure 3 c) of the main manuscript for healthy RBCs considering only membrane positions where optical traps were applied (orange data). Second column: average value of $\sigma_{\delta h}$, with uncertainty given by its standard deviation, evaluated for each sample at different trapping laser power $P_1$ and $P_2$. Right box: probability ($p$-value) calculated from pairwise $t$-test between $\sigma_{\delta h}$ evaluated for all membrane positions and $\sigma_{\delta h}$ only at positions where optical traps were generated, at laser power $P_1$ and $P_2$, respectively.

|              |  | $\sigma_{\delta h}$ (nm) | p-value ($\sigma_{\delta h}$ all positions vs traps positions) | |
|--------------|--|--------------------------|---|---|
| $P_1$ (traps) |  | 21.9 ± 2.5              | $P_1$ | $3 \times 10^{-2}$ |
| $P_2$ (traps) |  | 22.2 ± 2.0              | $P_2$ | $5 \times 10^{-3}$ |

**Table T5**. Analysis of the data reported in Figure 4 of the main manuscript for RBCs considering all tracked membrane positions under LatA (red data) and ATP- (blue data) treatments. Average value of $\sigma_{\delta h}$, with uncertainty given by its standard deviation, evaluated for each sample at different trapping laser power and the probability ($p$-value) calculated from every pairwise $t$-test.

| LatA | $\sigma_{\delta h}$ (nm) | $p$-value | | |
|---|---|---|---|---|
| | | $P_0$ | $P_1$ | $P_2$ |
| $P_0$ | 23.8 ± 2.5 | | | |
| $P_1$ | 21.9 ± 2.1 | $10^{-39}$ | | |
| $P_2$ | 21.8 ± 1.7 | $10^{-35}$ | 0.37 | |

| ATP- | $\sigma_{\delta h}$ (nm) | $p$-value | | |
|---|---|---|---|---|
| | | $P_0$ | $P_1$ | $P_2$ |
| $P_0$ | 16.7 ± 2.2 | | | |
| $P_1$ | 16.9 ± 2.2 | 0.22 | | |
| $P_2$ | 17.5 ± 2.4 | $10^{-7}$ | $10^{-5}$ | |

**Table T6**. Analysis of the data reported in Figure 4 of the main manuscript for LatA and ATP- treated RBCs considering only membrane positions where optical traps were applied (orange data). In the third column are reported the average value of $\sigma_{\delta h}$, with uncertainty given by its standard deviation, evaluated for each sample at different trapping laser power $P_1$ and $P_2$. In the right box are shown the probabilities ($p$-value) calculated from pairwise $t$-test between $\sigma_{\delta h}$ evaluated for all membrane positions and $\sigma_{\delta h}$ only at positions where optical traps were generated, for both samples at laser power $P_1$ and $P_2$, respectively.

| | | $\sigma_{\delta h}$ (nm) | | $p$-value ($\sigma_{\delta h}$ all positions vs traps positions) |
|---|---|---|---|---|
| LatA | $P_1$ (traps) | 23.1 ± 1.8 | $P_1$ | $4 \times 10^{-6}$ |
| | $P_2$ (traps) | 22.6 ± 1.7 | $P_2$ | $9 \times 10^{-3}$ |
| ATP- | $P_1$ (traps) | 17.7 ± 2.4 | $P_1$ | $3 \times 10^{-3}$ |
| | $P_2$ (traps) | 18.2 ± 2.5 | $P_2$ | $2 \times 10^{-2}$ |

**Table T7**. Average value of $\sigma_{\delta h}$ and $\sigma_{\delta F}$, with uncertainty given by its standard deviation, evaluated for each sample at laser power P$_1$ and P$_2$, corresponding to the force data reported in in the dynamic maps of Figure 5 of the main manuscript and Figure S11 of the Supplementary information, respectively.

|  | Power P$_1$ | | Power P$_2$ | |
| --- | --- | --- | --- | --- |
|  | $\sigma_{\delta h}$ (nm) | $\sigma_{\delta F}$ (fN) | $\sigma_{\delta h}$ (nm) | $\sigma_{\delta F}$ (fN) |
| Healthy | 21.9 ± 2.5 | 10 ± 3 | 22.1 ± 1.8 | 15 ± 5 |
| LatA | 23.0 ± 1.8 | 21 ± 5 | 22.3 ± 1.7 | 29 ± 6 |
| ATP- | 17.5 ± 2.4 | 11 ± 2 | 18.0 ± 2.4 | 15 ± 3 |
| Fixed | 5.3 ± 1.4 | 12 ± 4 | 5.9 ± 1.9 | 16 ± 5 |

**Table T8**. **Statistical analysis of the force data reported in Figure 5.** Second column: average value of $\sigma_{\delta F}$, with uncertainty given by its standard deviation, evaluated for each sample under optical trapping at laser power P$_1$. Right box: probability ($p$-value) calculated from all the pairwise $t$-tests.

|  | $\sigma_{\delta F}$ (fN) |  | $p$-value | | | |
| --- | --- | --- | --- | --- | --- | --- |
|  |  |  | Healthy | LatA | ATP- | Fixed |
| Healthy | 10 ± 3 |  |  |  |  |  |
| LatA | 21 ± 5 |  | < 10$^{-33}$ |  |  |  |
| ATP- | 11 ± 2 |  | 0.21 | < 10$^{-100}$ |  |  |
| Fixed | 12 ± 4 |  | 0.004 | < 10$^{-34}$ | 0.008 |  |

**Table T9**. Analysis of the data reported in Figure 6 b) of the main manuscript. Second column: average value of the dissipated power, with uncertainty given by its standard deviation, evaluated for each sample. Right panel: probability ($p$-value) calculated from all the pairwise $t$-tests of the power distributions.

|  | Power ($k_BT/s$) | $p$-value | | | |
|---|---|---|---|---|---|
|  |  | Healthy | LatA | ATP- | Fixed |
| Healthy | 18.1 ± 6.1 |  |  |  |  |
| LatA | 24.7 ± 8.5 | $10^{-7}$ |  |  |  |
| ATP- | 12.9 ± 6.3 | $10^{-5}$ | $10^{-14}$ |  |  |
| Fixed | 5.3 ± 2.9 | $10^{-14}$ | $10^{-15}$ | $10^{-7}$ |  |

**Table T10. Flickering variability of RBCs from 5 different healthy donors.** The experimental conditions for the controls were 25 °C. Values of $\sigma_{\delta h}$ represent mean ± SD across N=10 cells per donor. The flickering amplitude $\sigma_{\delta h}$ values were consistent across donors, with no systematic dependence on sex, age, or blood type. Inter-donor variability was found within ~10%, confirming reproducibility across healthy individuals. The donor listed as Control 4 in the table was used as the source of RBCs analyzed in the main manuscript.

| **Control ID** | **Sex / Age (years)** | **Blood serotype** | $\sigma_{\delta h}$ (nm) | N (cells) |
|---|---|---|---|---|
| Control 1 | Female / 17 | O− | 24 ± 4 | 10 |
| Control 2 | Male / 16 | A+ | 23 ± 3 | 10 |
| Control 3 | Female / 25 | B− | 25 ± 4 | 10 |
| Control 4 | Male / 23 | O+ | 21 ± 3 | 10 |
| Control 5 | Male / 45 | O+ | 22 ± 5 | 10 |